\documentclass[aps,prd,reprint,superscriptaddress,showpacs,preprintnumbers,amsmath,amssymb,nofootinbib]{revtex4-1}
\usepackage{graphicx}% Include figure files
\usepackage{dcolumn}% Align table columns on decimal point
\usepackage{amsfonts,amsthm,stmaryrd,mathtools,mathrsfs,bm}%bm  bold math
\usepackage{hyperref}% add hypertext capabilities
\usepackage{xcolor}
\usepackage{tikz}
\usepackage{tikz-cd}

\usepackage{cases}
\usepackage{subcaption}

\usepackage{listings}
\usepackage{array}
\usepackage{multirow}

\def\be{\begin{equation}}
\def\ee{\end{equation}}
\def\ba{\begin{eqnarray}}
\def\ea{\end{eqnarray}}

\newtheorem{theorem}{Theorem}

\newtheorem{corollary}[theorem]{Corollary}

\newtheorem{proposition}[theorem]{Proposition}
\theoremstyle{definition}
\newtheorem{remark}[theorem]{Remark}

\newcommand{\blue}{{}}

\newcommand\nn{\nonumber}
\newcommand{\q}{\quad}

\newcolumntype{L}[1]{>{\raggedright\let\newline\\\arraybackslash\hspace{0pt}}m{#1}}
\newcolumntype{C}[1]{>{\centering\let\newline\\\arraybackslash\hspace{0pt}}m{#1}}
\newcolumntype{R}[1]{>{\raggedleft\let\newline\\\arraybackslash\hspace{0pt}}m{#1}}

\newcommand{\intertwiner}[1]{
\begin{tikzpicture}[scale=0.2,rotate = #1]
\node[rectangle, fill, inner sep=1.8pt,rotate=#1] (v) at (0,0) {};
\node (n1) at (-2,2) {} ;
\node (n2) at (-0.6,2) {};
\node (n3) at (0.6,2) {};
\node (n4) at (2,2) {};

\draw [-{Circle[open]}, line width =0.7] (v) to [bend left] ++(n1) ;
\draw [-{Circle[open]}, line width =0.7] (v) to [bend left] ++(n2) ;
\draw [-{Circle[open]}, line width =0.7] (v) to [bend right] ++(n3) ;
\draw [-{Circle[open]}, line width =0.7] (v) to [bend right] ++(n4) ;
\end{tikzpicture}
}

\begin{document}

\title{Solving the area-length systems in discrete gravity using homotopy continuation }

\author{Seth K Asante}
\email[corresponding author: ]{seth.asante@uni-jena.de}
\affiliation{Theoretisch-Physikalisches Insit\"{u}t, Friedrich-Schiller-Universit\"{a}t Jena, Max-Wien Platz 1, 07743, Jena, Germany}

\author{Taylor Brysiewicz}
\email{tbrysiew@uwo.ca}
\affiliation{Department of Mathematics, University of Western Ontario, 2004 Perth Dr, London, ON N6G 2V4, Canada}

\begin{abstract}

Area variables are intrinsic to connection formulations of general relativity, in contrast to the fundamental length variables prevalent in metric formulations. Within 4D discrete gravity, particularly based on triangulations, the area-length system establishes a relationship between area variables associated with triangles and the edge length variables. This system is comprised of  polynomial equations derived from Heron's formula, which relates the area of a triangle to its edge lengths. 

Using tools from numerical algebraic geometry, we study the area-length systems. In particular, we show that given the ten triangular areas of a single $4$-simplex, there could be up to $64$ compatible sets of edge lengths. Moreover, we show that these $64$ solutions do not, in general, admit formulae in terms of the areas by analyzing the Galois group, or monodromy group, of the problem. We show that by introducing additional symmetry constraints, it is possible to obtain such formulae for the edge lengths. We take the first steps toward applying our results within discrete quantum gravity, specifically for effective spin foam models.

\end{abstract}
\maketitle
%\tableofcontents

\section{Introduction}

The fundamental framework of classical general relativity (GR) is built upon a geometric description of spacetime using metric variables that quantify the lengths of curves within a spacetime manifold. Alternatively, in several approaches to quantum gravity, {\it area variables} naturally appear as their fundamental degrees of freedom. 
This is particularly prominent in approaches based on connection formulations of GR, such as loop quantum gravity \cite{Ashtekar:1986yd} and spin foam models \cite{Perez:2012wv}.  In these approaches, area variables encode quantum geometric data  \cite{Engle:2007uq}. Interestingly, this notion resonates across diverse fields: Holography incorporates area variables, particularly in the reconstruction of geometry from entanglement \cite{Ryu:2006bv,VanRaamsdonk:2010pw}. 
Furthermore, in black hole physics \cite{Bekenstein:2015soa,Ashtekar:1997yu,BarberoG:2015xcq}, the entropy of a black hole is tied to the area of its event horizon through the Bekenstein-Hawking formula \cite{Bekenstein:1973ur}. 

Area variables are fundamentally distinct from length or metric variables, they offer a generalized notion of geometries compared to traditional length geometries \cite{Schuller:2005yt}. Within loop quantum gravity (LQG) and spin foam models, area variables lead to an extended configuration space of discrete geometries \cite{Dittrich:2008ar,Dittrich:2010ey,Freidel:2010aq}. The extended configuration space, also inherent in topological theories, facilitates an exact quantization of these discrete geometries. Moreover, both canonical \cite{Asante:2018wqy} and perturbative continuum limit analysis \cite{Dittrich:2021kzs,Dittrich:2022yoo} in the discrete gravity setting suggests that area variables possess more degrees of freedom than their length counterparts. Establishing a relationship between geometries described by area variables and those described by length geometries is essential in bridging the gravitational approaches based on area variables with metric gravity. This article concentrates on discrete geometries based on triangulations to provide such a relationship.

Several discrete quantum gravity approaches are based on a piecewise flat or piecewise linear approximation of spacetime manifolds as regulators, often represented by triangulations.
Within such a triangulation, an {\it area geometry} is an assignment of area values to its two-dimensional faces (triangles). The classical gravitational dynamics for a given area geometry on a fixed triangulation is aptly described by {\it area Regge calculus} \cite{Makela:2000ej,Barrett:1997tx,Asante:2018wqy} akin to length Regge calculus \cite{Regge:1961px} for length geometries.  
The connection between area geometries and loop quantum gravity, initially explored in \cite{Rovelli:1993kc} also features in the construction and semi-classical analysis of vertex amplitudes in spin foam models \cite{Barrett:1997gw,Conrady:2008mk,Hellmann:2013gva,CamoesdeOliveira:2016rcm}.
Effective spin foam models \cite{Asante:2020qpa,Asante:2021zzh,Asante:2020iwm}, recently constructed as path integral models for four dimensional discrete quantum gravity, are also based on triangulations. These models directly utilize {\it discrete area variables} assigned to triangles as their fundamental degrees of freedom to define transition amplitudes for the area geometries. 

Nonetheless, unlike length geometries, area geometries pose challenges in defining geometric quantities such as volumes and angles within a triangulation. This difficulty stems from inherent ambiguities in solely using areas of triangles to describe geometric quantities. For this reason, area Regge calculus encountered several ambiguities 
\cite{Barrett:1997tx}. This article  quantifies these ambiguities for generic triangulations.

Heron's formula provides a useful tool for relating the area and length variables within a triangulation. It expresses the area of a triangle in terms of its edge lengths:
\be\label{Heron}
A = \frac{1}{4}\sqrt{4\ell_1^2\ell_2^2 - (\ell_1^2+\ell_2^2-\ell_3^2)^2 }.
\ee
By considering a set of areas associated with triangles in a triangulation,  Heron's formula can be used to explore relationships between the area and length geometries. Naturally, this formulation leads to a system of polynomial equations (see Equations \eqref{AlSys2} and \eqref{TrgsHeron})  referred to as an {\it area-length system}, where areas of triangles serves as parameters and edge lengths as variables.  Importantly, within spin foam models, the area-length system intertwines with the concept of {\it simplicity constraints} \cite{Engle:2007uq}. These constraints are imposed on discrete quantum geometries aiming to recover classical geometries in the appropriate limit.
  
For an arbitrary triangulation, the area-length system is complicated and closed form solutions are difficult to find. Even for a single $4$-simplex, solving the area-length system for generic areas is non-trivial. In this context, the field of numerical algebraic geometry provides valuable methods ideally suited to numerically solve various polynomial systems of equations.  Specifically, we show how to use {\it homotopy continuation methods} (see \cite[Section 2]{NumericalNonlinearAlgebra}) for solving these systems. 

Ideally, one seeks an explicit general formula for the ten edge lengths of a 4-simplex given its ten triangular areas. We show that this is impossible for two reasons. The first is that given ten triangular areas of a $4$-simplex, there can be up to $64$ possibilities for the corresponding $10$-tuples of edge lengths (see Theorem \ref{thm:64solutionsnumerical}$^*$). The next hope would be to have some formula in terms of the areas, involving standard operations like  addition, subtraction, multiplication, division, and $n$-th roots, for each of these $64$ (complex) solutions. By  analyzing the monodromy group (or Galois group) of this problem (Theorem \ref{thm:monodromygroup}), we show that no such formula exists (Corollary \ref{cor:notsolvable}). However, by restricting the area parameters to have special symmetries, the corresponding Galois group becomes smaller and potentially solvable. We showcase this phenomenon in the most symmetric setting when all of the areas of a $4$-simplex are the same. There, the Galois group is solvable, and the $64$ solutions are expressible in formulae involving basic operations and square roots.
For a more general treatment of these Galois groups in the context of volumes of simplices, see \cite{Heron}. The non-existence of such a formula underlines the importance of using numerical techniques like homotopy continuation to locally solve for and track the solutions to the area-length system.

\vspace{0.4em}

\noindent
{\bf  Outline of paper:} In Section \ref{sec:DiscG}, we describe the area-length system for a 4-simplex and apply numerical homotopy continuation methods, discussed in Section \ref{sec:Homotopy}, to obtain isolated solutions to this system. A succinct algorithm to solve the area-length system for a 4-simplex is provided in Appendix \ref{app:Jcode}. In Section \ref{sec:Results}, we collect the results of a computational experiment where we solve $3$ million instances of the area-length system of a $4$-simplex numerically, under various distributions on the area parameters. We summarize the behavior of the $64$ solutions under these distributions in terms of the number of real solutions, positive solutions, and the geometric nature of the solutions. Additionally, we give area parameters for which there exist $64$ real solutions, half of which correspond to Lorentzian simplices.
 In Section \ref{sec:symmetry}, we outline how including symmetry constraints on the area parameters or edge variables influences the solutions to the area-length system, paying particular attention to when all areas are equal. 
We show that such restrictions can change the solvability of the area-length system.  We extend the homotopy methods to address area-length systems for general triangulations in Section \ref{sec:Triangulations}. The paper concludes in Section \ref{sec:App} with a discussion on the implications of our findings within discrete gravity and effective spin foam models.

\section{ Discrete Geometries: Length and Area Variables } \label{sec:DiscG}

In discrete gravity, based on triangulations, geometric quantities are assigned to components or subsets of the simplices within the triangulation. The choice of which geometric quantities are considered fundamental varies across different approaches. 
Regge calculus \cite{Regge:1961px}, for instance, is conceptualized as a discretization for general relativity based on triangulations.  In its original formulation, lengths assigned to the edges of the triangulation are considered as fundamental variables. Each simplex is taken to be flat and equipped with a {\blue length geometry}, where the edge lengths are considered fundamental. These simplices are then glued together into a triangulation by matching the length geometries across shared sub-simplices. A rigorous quantization of discrete geometries starting from these length variables remains an open issue. 

In four dimensions, area Regge calculus provides an alternative description wherein areas associated to triangles are considered fundamental. The assignment of areas to the triangles of the triangulation defines its {\blue area geometry}. Here, the 4-simplices are equipped with area geometries and then glued together by ensuring that the areas of triangles across shared sub-simplices match. 
These area geometries within triangulations offer more flexibility compared to length geometries, allowing for configurations that do not necessarily require ``shape-matching'' across shared tetrahedra. This feature makes area geometries particularly relevant within spin foam models for quantum gravity. Effective spin foam models provide a quantization of discrete geometries, directly utilizing {\it discrete} area variables within a triangulation as fundamental. 

Heron's formula \eqref{Heron}  relates the area of a triangle to its edge lengths. When applied to all triangles in a triangulation, it  establishes a relation between area geometry and possible length geometries. The set of equations, obtained by applying Heron's formula to each  triangle, constitutes the {\textit{area-length system}} associated to that triangulation. The inversion of areas for lengths through the area-length system is useful in defining geometric quantities associated to simplices. It is also a crucial step in the semi-classical analysis of spin foam models \cite{Han:2021kll} which relates to Regge calculus. 
Given the areas of triangles, considered as parameters, the solution set to an area-length system yields a collection of edge lengths that are compatible with the area geometry for the triangulation. In this article, we explore the sets of length solutions to area-length systems within triangulations and examine their implications.

\subsection{ Area-length system for a 4-simplex}

We begin our exploration with the simplest triangulation in four dimensions: a single 4-simplex. A geometric flat 4-simplex, denoted by $\blue \Delta_4$, is determined by five points in $\mathbb R^{4}$ (not all on a hyperplane)  called the vertices of the $4$-simplex. 
The edges of $\Delta_4$ are the $\binom{5}{2}=10$ line segments formed by connecting pairs of vertices. All geometric quantities pertaining to $\Delta_4$, like volumes and angles, can be computed from these edge lengths. Notably, the squared volumes of all faces of the simplex are scaled principle minors of a Euclidean distance matrix (see \cite{Heron}). Such minors are called \textit{Cayley-Menger determinants}.

Beyond its ten edges, a $4$-simplex $\Delta_4$ has $\binom{5}{3} = 10$ triangular faces, each containing three of its vertices. Although the number of triangles of $\Delta_4$ equals the number of edges, it is well-known that the triangular areas of $\Delta_4$ do not, in general, determine its geometry \cite{Barrett:1997tx}. 

In fact, the same set of areas of triangles of a $4$-simplex may correspond to multiple edge length assignments. 
An illustrative example is given by the following: Consider two $4$-simplices $\Delta_4$ and $\Delta_4'$, with edge lengths $\{\ell_{ij}\}_{1 \leq i<j \leq 5}$ and $\{\ell'_{ij}\}_{1 \leq i<j \leq 5}$, respectively. Taking~these edge lengths to be
\begin{numcases}{}
\ell_{12} = 2+ 2\sqrt{1-4\alpha^2} , \q \ell_{ij} =1 \text{ for all others} \nn \\
\ell'_{12} = 2- 2\sqrt{1-4\alpha^2} , \q \ell'_{ij} =1 \text{ for all others} 
\end{numcases} 
for some $\alpha \in (0,\tfrac12)$.
These two $4$-simplices  share an identical area geometry: the triangle areas are $A'_{12i} =A_{12i} = \alpha$ for $i=3,4,5$ and all remaining triangles have area $\sqrt{3}/4$. 

The area geometry of a 4-simplex $\Delta_4$  is described by assigning area values to its triangles. Writing Heron's formula for each of the ten triangles connects the area geometry and the length geometry of the 4-simplex through the following ten equations:
\begin{numcases}{}\label{AlSys}
16 A_{123}^2 =  {4 \ell_{12}^2 \ell_{13}^2 - (\ell_{12}^2+\ell_{13}^2 - \ell_{23}^2)^2 } \nn \\
16 A_{124}^2 =  {4 \ell_{12}^2 \ell_{14}^2 - (\ell_{12}^2+\ell_{14}^2 - \ell_{24}^2)^2 } \nn \\
\q \q \vdots    \q  \,\, \vdots \q \q  \q \vdots 
\\
16 A_{345}^2 =  {4 \ell_{34}^2 \ell_{35}^2 - (\ell_{34}^2+\ell_{35}^2 - \ell_{45}^2)^2 }  \nn \q \q 
\end{numcases}
Here, ${\blue{A_{ijk}}}$ is the area parameter of the triangle labelled by the vertices $i,j,k$ and ${\blue{\ell_{ij},\ell_{ik},\ell_{jk}}}$ are its edges lengths. These ten polynomial equations constitute the area-length system (or the Heron's system) for the 4-simplex.  

\section{ Homotopy Continuation: Exploring solutions numerically }\label{sec:Homotopy}

As a consequence of its nonlinear features, the area-length system \eqref{AlSys} admits many solutions given fixed areas. Popularized by the emerging field of numerical algebraic geometry, the numerical method of \textit{homotopy continuation} is a tool for reliably computing floating-point approximations of solutions to polynomial systems, like \eqref{AlSys}. For detailed background on numerical algebraic geometry, we invite the reader to consult \cite{IntroNumericalAlgGeom:2005,NumericalNonlinearAlgebra}. 

{ It is important to point-out that the computational tools outlined in the subsequent sections work over the complex numbers $\mathbb{C}$. A main reason for this is that many results can be more uniformly stated for polynomial systems over the complex numbers. For example, the fundamental theorem of algebra states that any univariate polynomial of degree $n>0$ has $n$ complex roots, counted with their appropriate multiplicities, but the number of real roots varies depending on the coefficients. Similarly, for the area-length systems considered in this paper, we enjoy the conclusion of the \textit{parameter continuation theorem} (see \cite{ParameterContinuation, IntroNumericalAlgGeom:2005}). We follow the standard method for gleaning real or semi-algebraic information about the solutions to a polynomial system: first solve it over $\mathbb{C}$ and then post-process those solutions (e.g. count how many are real, positive, etc). }

\subsection{Background on homotopy continuation} 

\begin{figure}[!htpb]
\includegraphics[scale=0.3]{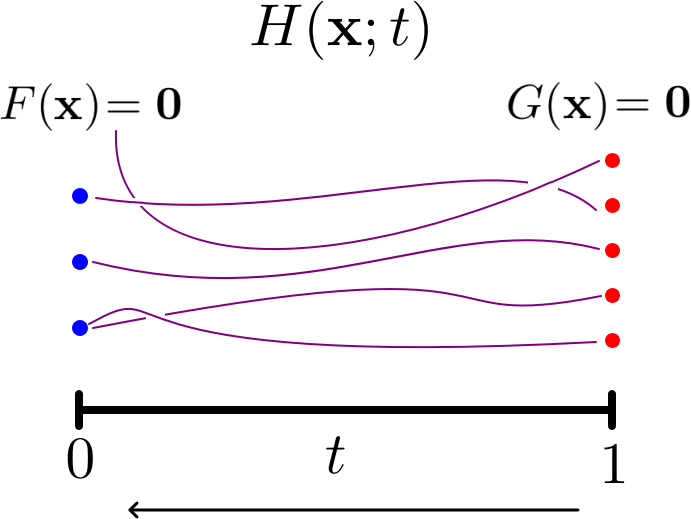}
\caption{An illustration of the numerical method of homotopy continuation.}
\label{fig:homotopy}
\end{figure}

Consider a collection ${\blue{F}}=\{{\blue{f_1,\ldots,f_n}}\}$ of ${\blue{n}}$ polynomials in $n$ variables ${\blue{\bf x}} = ({\blue{x_1,\ldots,x_n}})$, with coefficients in the field $\mathbb{C}$ of complex numbers. Such a system is called {\blue{square}} since the number of equations in $F$ equals the number of variables. The goal of {\blue{homotopy continuation}} is to compute the numerical approximations of isolated solutions to this square system, called the  {\blue{target system}}:
\be
F(\textbf{x}) = \begin{bmatrix} f_1(\textbf{x}) \\ \vdots \\ f_n(\textbf{x}) \end{bmatrix} = \textbf{0}.
\ee
The main idea is to construct a simpler polynomial system ${\blue{G(\textbf{x})=0}}$, called the {\blue{start system}}, which is similar to $F(\textbf{x})=0$, but is easier to solve.
A standard choice for such a start system is the {\blue{total degree}} start system:
\be
\label{totaldegree}
{\blue{G_{\textrm{total}}(\textbf{x})}} = \begin{bmatrix} g_1(\textbf{x}) \\ \vdots \\ g_n(\textbf{x})\end{bmatrix} = 
\begin{bmatrix} x_1^{\textrm{deg}(f_1)}-1 \\ \vdots \\ x_n^{\textrm{deg}(f_n)}-1 \end{bmatrix}.
\ee
The $\prod_{i=1}^n \textrm{deg}(f_i)$-many solutions  to $G_{\textrm{total}}(\textbf{x})=\textbf{0}$, called the {\blue start solutions}, are all trivial to find. 
 The number  $\prod_{i=1}^n \textrm{deg}(f_i)$,  called the {\blue{B\'ezout bound}} of the system $F(\textbf{x})={\bf 0}$, is an upper bound for the number of its isolated solutions.  Next, one constructs a {\blue{homotopy}}
\be
\label{totaldegreehomotopy}
{\blue{H(\textbf{x};t)}} = (1-t)F(\textbf{x}) + \xi \cdot t\cdot G_{\textrm{total}}(\textbf{x}) \quad \quad \xi \in \mathbb{C}
\ee
that interpolates between the start and the target system as $t$ goes from $1$ to $0$. Here, ${\blue{\xi}}$ is a random complex number.  The known solutions to the start system are analytically continued along solution paths over $t$ via standard numerical predictor-corrector methods as illustrated in Figure \ref{fig:homotopy}. The complex number $\xi$, chosen at random, ensures (with probability one) that the solution paths do not cross each other. Numerical  {\blue{endgames}} are employed near $t=0$ to obtain the solutions to $F(\textbf{x})={\bf 0}$. 

The process of homotopy continuation, applied to \eqref{totaldegreehomotopy}, finds all isolated solutions to $F(\textbf{x})=\textbf{0}$. Moreover, this technique can be applied even to non-square systems by first ``squaring-up'' the system. We, again, encourage the reader to consult \cite{IntroNumericalAlgGeom:2005, NumericalNonlinearAlgebra} for details.

\subsection{Precomputation: solving the area-length system once}
To utilize homotopy continuation methods for the area-length system, we introduce the variables ${\blue{x_{ij}}} = \ell_{ij}^2$, representing the squared edge lengths and ${\blue{v_{ijk}}} = A_{ijk}^2$, representing the squared triangular areas. This change of variables allows us to cast Heron's formula \eqref{Heron} as a polynomial equation 
\begin{equation}\label{HeronSq}
16v_{ijk}= \hspace{-1pt} 4 x_{ij}x_{ik} \hspace{-1pt}-\hspace{-1pt} (x_{ij} + x_{ik} - x_{jk})^2 
\end{equation}
The area-length system \eqref{AlSys} becomes a system of ten polynomials in ten variables $\textbf{x}=(x_{12},\ldots,x_{45})$ and ten parameters $\textbf{v} = (v_{123},\ldots,v_{345})$:

\begin{align} \label{AlSys2}
\hspace{-5pt}{\blue{F_{\Delta_4}(\textbf{x};\textbf{v})}} &={\footnotesize{ \begin{bmatrix}
    &4 x_{12} x_{13} - (x_{12}+x_{13} -x_{23})^2 - 16 v_{123} &\vspace{2ex}\\ 
    &4 x_{12} x_{14} - (x_{12}+x_{14} -x_{24})^2 - 16 v_{124} &\vspace{1ex} \\ 
    &\vdots \vspace{1ex}\\ 
    &4 x_{34} x_{35} - (x_{34}+x_{35} -x_{45})^2 - 16 v_{345}&
\end{bmatrix} }}
\end{align}
Henceforth, \eqref{AlSys2} is what we refer to when we discuss the {\blue{area-length system}} for the simplex $\Delta_4$.

Now written as a square polynomial system, an instance of the area-length system is easily solved by the method of homotopy continuation: we simply choose random values in the parameter space $\textbf{v}^* \in \blue{\mathbb{C}_{\textbf{v}}^{10}}$  and perform homotopy continuation from the total degree start system \eqref{totaldegree} to the system $F_{\Delta_4}(\textbf{x};\textbf{v}^*)$ along the homotopy \eqref{totaldegreehomotopy}. This computation takes no more than a few seconds in any standard piece of numerical algebraic geometry software (e.g. the \texttt{julia} package \texttt{HomotopyContinuation.jl} \cite{HCjl,julia}). It tracks a total of $2^{10}=1024$-many start solutions and produces $64$ target solutions in $\blue{\mathbb{C}_{\textbf{x}}^{10}}$. The $960$ excess paths approach $\infty$ as $t$ approaches $0$ as depicted in Figure \ref{fig:homotopy}. The following theorem comes with a $*$ symbol, as is standard in the field of numerical algebraic geometry, to indicate that the statement is a result of numerical computation and hence is subject to numerical error.
\begin{theorem}[*]\label{Thm:Solutions}
\label{thm:64solutionsnumerical}
Given generic areas $\textbf{v} \in \mathbb{C}_{\textbf{v}}^{10}$, the area-length system \eqref{AlSys2} has $64$ isolated solutions in~$\mathbb{C}_{\textbf{x}}^{10}$.
\end{theorem}
{ Here, \textit{generic} means \textit{outside a set of measure zero} in $\mathbb{C}_{\textbf{v}}^{10}$, called the \textit{discriminant} of the problem. }

\subsection{A parameter homotopy: solving the area-length system fast}
The family $F_{\Delta_4}({\textbf x}; {\bf v})={\bf 0}$ of polynomial systems, parametrized by squared triangular areas, fits the hypotheses of the {\blue{Parameter Continuation Theorem}} (see \cite{ParameterContinuation, IntroNumericalAlgGeom:2005}). Briefly, it  states that for almost all choices of ${\bf v}^* \in \mathbb{C}_{\bf v}^{10}$, there is some number ${\blue{N}}$ of isolated solutions to $F_{\Delta_4}({\bf x}; {\bf v}^*)={\bf 0}$ in $\mathbb{C}_{\textbf{x}}^{10}$ and that this  number is maximal among all parameters. Those parameters for which there are not $N$ solutions form a subset ${\blue{\textrm{Disc}}}$ of $\mathbb{C}_{\bf v}^{10}$ of measure zero which we call the {\blue{discriminant}}. The parameters not in the discriminant are called {\blue{generic}}.  Theorem~\ref{thm:64solutionsnumerical}$^{*}$ claims that $N=64$. Using certification methods in \texttt{HomotopyContinuation.jl} relying upon interval arithmetic, we show that there exists a system $F({\bf x};\textbf{v}^*)=\textbf{0}$ with $64$ isolated solutions. This \textit{proves} that $N \geq 64$. See \cite{Certify} for details about this certification process. 

With this in mind, after an initial computation of the $64$ solutions to $F_{\Delta_4}(\bf x, \bf v^{(1)})$ for some generic ${\bf v^{(1)}} \in \mathbb{C}_{\bf v}^{10}$, one may construct a {\blue{parameter homotopy}}  to find the solutions to some other parameter value ${\bf v^{(2)}} \in   \mathbb{C}_{\bf v}^{10}$: 
\be
\label{eq:parameterhomotopy}
H({\bf x}; t) = F_{\Delta_4}({\bf x}; (1-t){\bf v^{(2)}}+t{\bf v^{(1)}})
\ee
Performing the path-tracking procedure over \eqref{eq:parameterhomotopy} is significantly cheaper than the initial computation of the solutions over ${\bf v^{(1)}}$ via a total degree homotopy \eqref{totaldegreehomotopy}: instead of tracking $1024$ paths, one need only track $N=64$. 

Systems which satisfy the parameter continuation theorem are also sometimes called {\blue{enumerative problems}}. The new software \texttt{Pandora.jl} \cite{Pandora}, built on top of \texttt{HomotopyContinuation.jl} \cite{HCjl} and \texttt{OSCAR} \cite{OSCAR}, is designed to automatically generate experimental data about enumerative problems via parameter homotopy computations. In the following section, we summarize this experimental data for the area-length system.

\section{Real, positive, and geometric solutions}\label{sec:Results}
Although Theorem \ref{thm:64solutionsnumerical}$^*$ suggests that there could be up to $64$ different $4$-simplices which exhibit a given area geometry, this is never the case because many of the solutions to the area-length system are non-real, have negative coordinates, or do not satisfy geometric inequalities like generalizations of the triangle inequality (see Proposition \ref{Realizable}). Using the \texttt{explore} and \texttt{optimize} features of  \texttt{Pandora.jl} \cite{Pandora} we generate data about the solutions to random area-length systems under various distributions.

\subsection{Real and positive length solutions}\label{sec:RPSols}

\begin{figure*}[ht!]
\centering
\begin{subfigure}[b]{0.3\textwidth}
\includegraphics[scale=0.3]{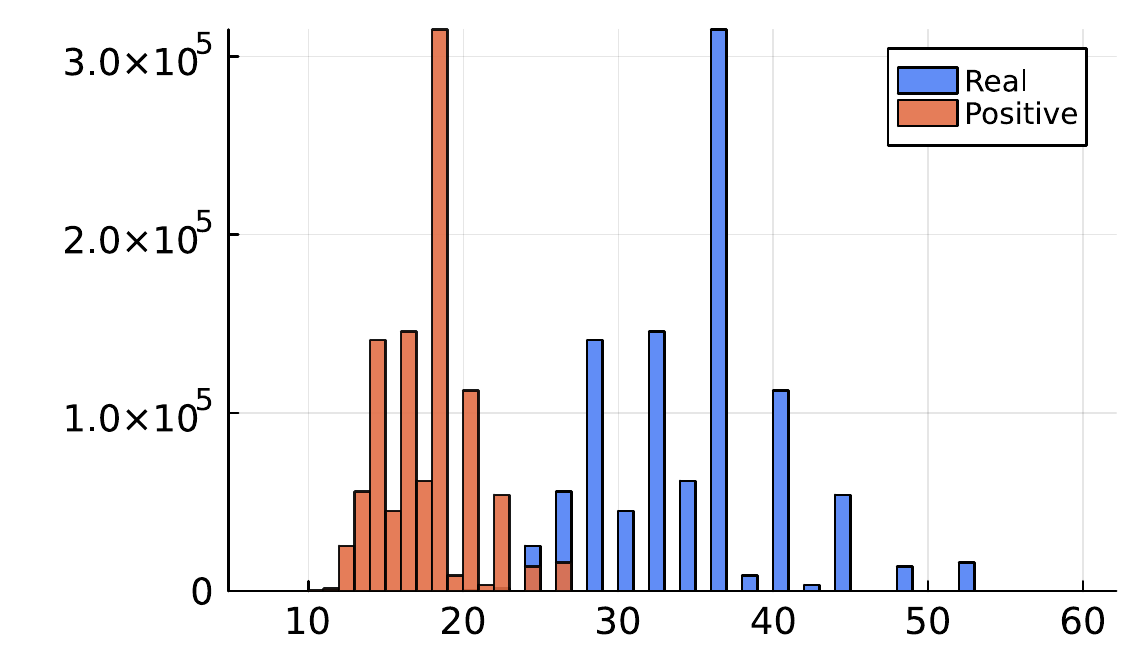}
\caption{Area parameters are randomly selected from the interval $I_{0.5}$}
\label{fig:histA}
\end{subfigure}
\hfill 
\begin{subfigure}[b]{0.3\textwidth}
\includegraphics[scale=0.3]{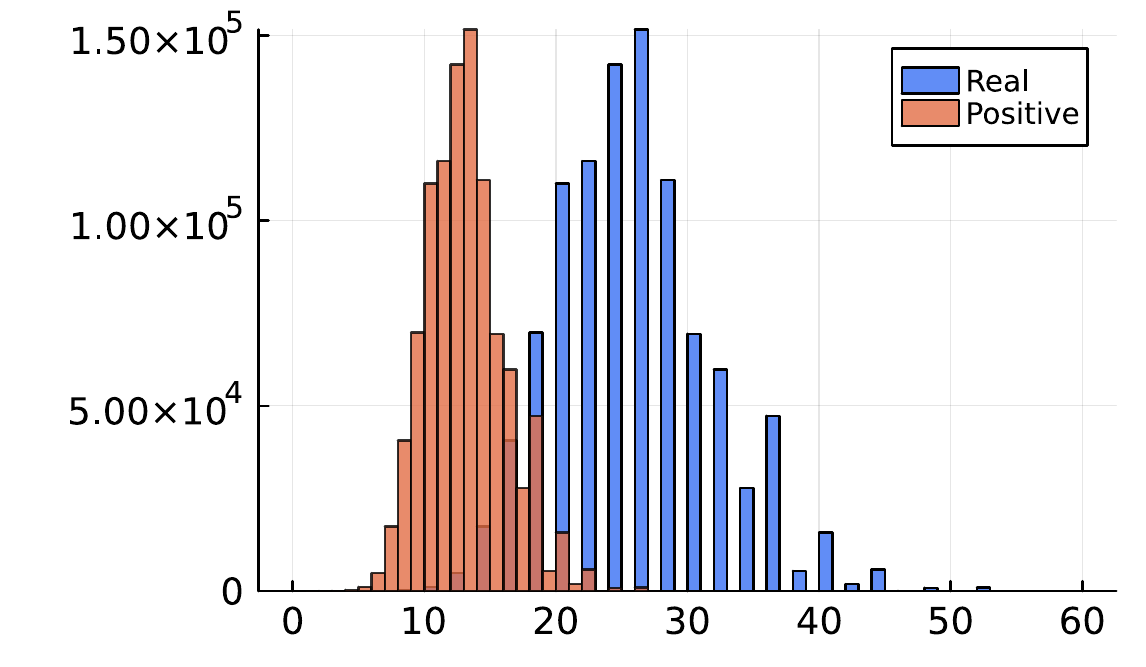}
\caption{Area parameters are randomly selected from the interval $I_{0.3}$}
\label{fig:histB}
\end{subfigure} \hfill 
\begin{subfigure}[b]{0.3\textwidth}
\includegraphics[scale=0.3]{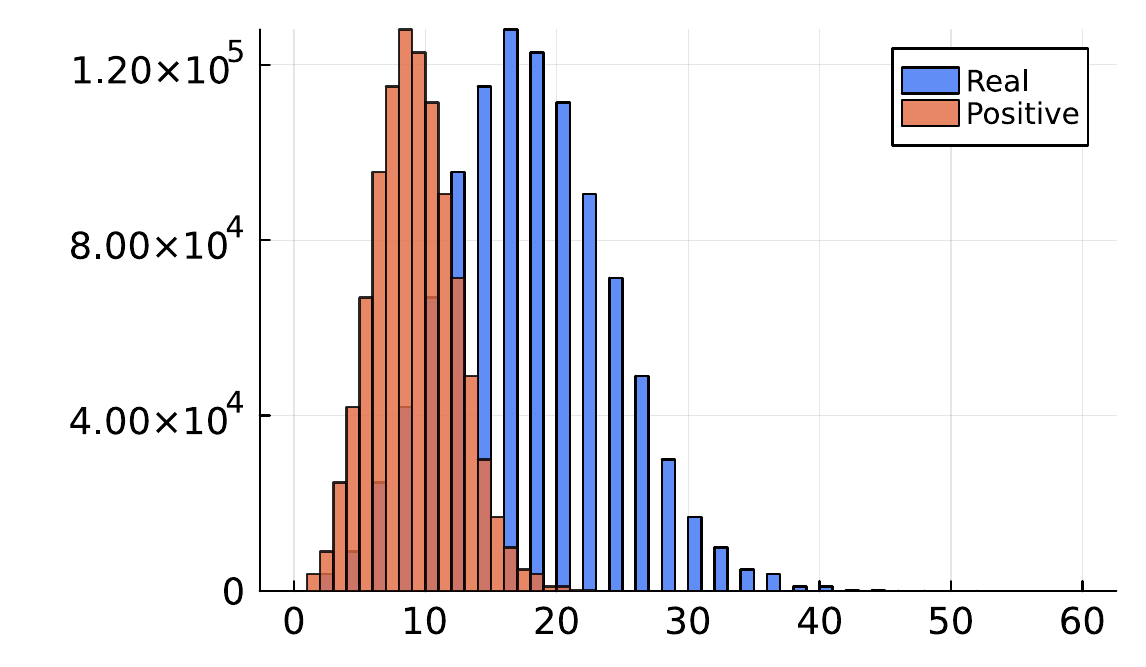}
\caption{Area parameters are randomly selected from the interval $I_{0.0}$}
\label{fig:histC}
\end{subfigure} \hfill 
\caption{Histograms for the counts of real and positive length solutions. The area parameters are randomly selected from three different intervals of $\mathbb R_{>0}$. Each histogram is generated from $\color{blue} 10^6$ samples of squared area parameter sets. }
\label{fig:hist}
\end{figure*}

\begin{figure*}[htb!]
\centering
\begin{subfigure}[b]{0.3\textwidth}
\includegraphics[scale=0.3]{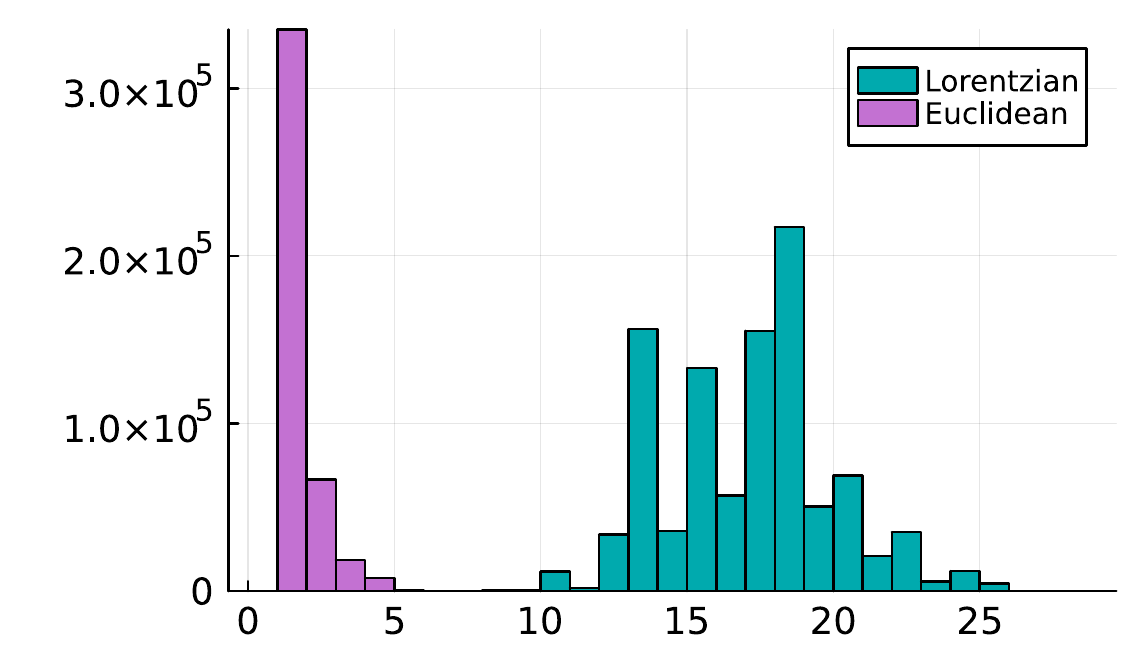}
\caption{Area parameters are randomly selected from the interval $I_{0.5}$}
\label{fig:hist2A}
\end{subfigure}
\hfill 
\begin{subfigure}[b]{0.3\textwidth}
\includegraphics[scale=0.3]{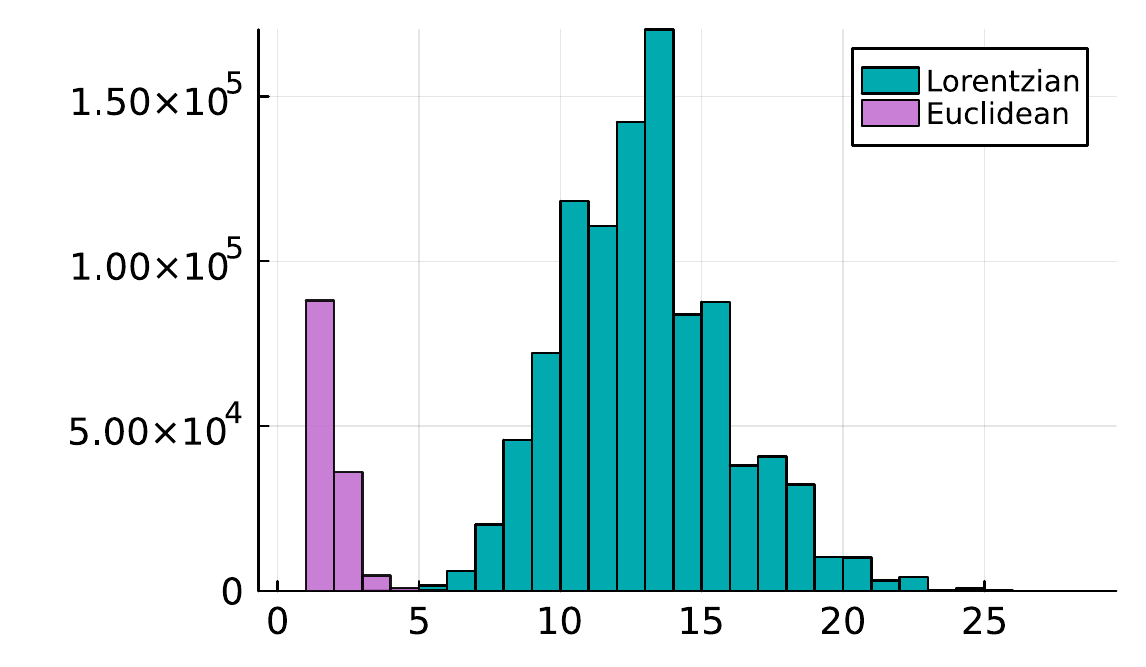}
\caption{Area parameters are randomly selected from the interval $I_{0.3}$}
\label{fig:hist2B}
\end{subfigure} \hfill 
\begin{subfigure}[b]{0.3\textwidth}
\includegraphics[scale=0.3]{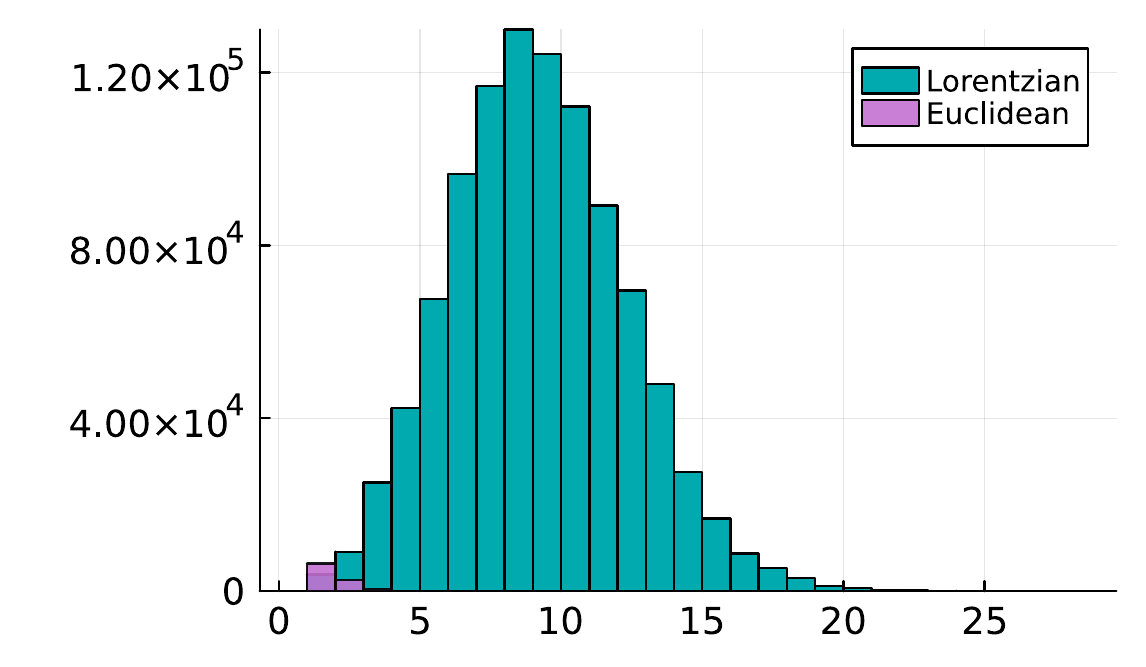}
\caption{Area parameters are randomly selected from the interval $I_{0.0}$}
\label{fig:hist2C}
\end{subfigure} \hfill 
\caption{Histograms for the counts of Euclidean and Lorentzian length solutions. The area parameters are randomly selected from three different intervals of $\mathbb R_{>0}$. Each histogram is generated from $\color{blue} 10^6$ samples of area parameter sets. }
\label{fig:hist2}
\end{figure*}

We concentrate on positive area parameters, ${\bf v}~\in~\mathbb R_{>0}^{10}$. Observe that over the real numbers $\mathbb R$, the squared area of a triangle is positive if and only if all its three edge lengths are either all positive or all negative. Thus, among all real solutions to \eqref{AlSys2} over positive area parameters ${\bf v}~\in~\mathbb R_{>0}^{10}$, half of them are positive and half are negative in the sense that they have all positive/negative coordinates respectively.

We consider multiple distributions in the parameter space $\mathbb{R}_{>0}^{10}$ of area parameters. Specifically, for any $\alpha \geq 0$, we consider distribution on ${\bf v} = (A_{123}^2,A_{124}^2,\dots,A_{345}^2)$ induced by choosing each non-squared area $A_{ijk}$ uniformly from the intervals $\blue{I_{\alpha}}=(\alpha,1)$.
In a computational experiment, we generated {\blue 1,000,000} sets of area parameters sampled uniformly at random from the interval $I_\alpha$ for $\alpha  \in \{0.0, 0.3, 0.5\}$. Subsequently, we solved the area-length system \eqref{AlSys2} for each set of parameters. The numbers of real and positive solutions found in these experiments are illustrated in the subfigures \ref{fig:histA}, \ref{fig:histB}, and \ref{fig:histC} of Figure \ref{fig:hist} and summarized in Table \ref{tab:Real_Sols}. We remark that as $\alpha \to 0.0$, the distribution of real and positive solutions seems to approach a Gaussian distribution.
\begin{table}[ht!]
\centering
\begin{tabular}{|C{4em}|C{7em}|C{3em}|C{3em}|C{3em}|}\hline
 Interval &  Mean/$64$ $=\%$&  Min  & Max & Mode \\ \hline
 $I_{0.0}$ &  $27.33\%$ & $0$ & $52$ & $16$  \\ \hline 
  $I_{0.3}$ &  $39.60\%$ & $4$ & $60$ & $26$  \\ \hline 
   $I_{0.5}$ &  $53.67\%$ & $20$ & $52$ & $36$  \\ \hline 
\end{tabular}
\caption{A Summary of Empirical Data on the Number of Real Length Solutions}
\label{tab:Real_Sols}
\end{table}

\textit{A priori} the number $64$, of complex solutions to the area-length system, bounds the number of real solutions. Using the \texttt{optimize} feature of \texttt{Pandora.jl} we were able to find area  parameters consisting of positive integers  for which the area-length system has \textit{all} $64$ solutions real. The area parameters are given by: 

\begin{align}\label{64sols}
& A_{123} = 1 , \q \q  \,\,\,\,\,\,\,\,\, A_{124} = 11251 ,  \q \,\,\,  A_{125} = 5823,  \nn\\
& A_{134} = 5804 , \q \q  A_{135} = 2852 ,  \q \q  A_{145} =  1,  \\
& A_{234} = 5350 , \q \q A_{235} = 2972,  \q \q  A_{245} = 16615 ,  \q \q \nn \\ 
& A_{345} = 19144 . \q \q \nn
\end{align}
Although just numerically solving the area-length system over these parameters does not provide a proof, due to the potential of  numerical errors, one can use the \texttt{certify} command in \texttt{HomotopyContinuation.jl} to  \textit{prove} that the corresponding area-length system has (at least) $64$ distinct isolated real solutions.  We certified that this system has $64$ real solutions even when $A_{145}$ is replaced with any integer $1 \leq {\blue P} \leq 207$. 

\begin{theorem}
\label{thm:64Real}
There exists integer areas so that there are $64$ real isolated solutions to the area-length system.
\end{theorem}
{ Notably, Theorem \ref{thm:64solutionsnumerical}, does not, \textit{a priori}, state that there can be $64$ \textit{physically relevant} solutions to the area-length system. The condition that a solution is ``geometric'' is a semi-algebraic condition depending on the signature of some associated matrix, as discussed in the next section.

\begin{remark}
\label{rem:rationals}
Theorem \ref{thm:64Real} was obtained by first finding floating-point area values which admit $64$ real solutions. We use three facts to turn these floating-point values into integer ones without changing the real structure of the solution set: 
\begin{enumerate}
\item The number of real solutions is an open condition. That is, given a generic parameter $A$ with $64$ distinct solutions, $k$ of which are real, there exists an open neighborhood of parameters around $A$ which \textit{all} have $k$ real solutions. 
\item The number of real-solutions to the area-length system is invariant under scaling the parameters since this corresponds to scaling the simplex. 
\item The rational numbers are dense in the reals.
\end{enumerate}
Thus, to turn our floating-point example into an integral one, we approximate our parameters by rational parameters until the real structure stabilizes, and then we scale the rational parameters by a constant factor to clear denominators.
\end{remark}
}

\subsection{Geometric and non-geometric solutions}

The edge lengths of a geometric simplex are subject to  the  {\blue simplex inequalities}, which generalize the 2D triangle inequality.  The simplex inequalities ensure that, given edge lengths, there exists an $n$-simplex in flat spacetime $\mathbb R^n$ with those edge lengths. For our purposes, we will focus on $4$-simplices that can be embedded in either Euclidean spacetime $\mathbb{R}^4$ or Minkowski spacetime $\mathbb{R}^{1,3}$. 

To describe the simplex inequalities of a 4-simplex $\Delta_4$ consider its $4\times 4$ Gram matrix ${\blue{M_{\Delta_4}}} = [m_{ij}]$ where 
\be\label{GramMatrix}
{\blue{m_{ij}}} = \tfrac12 \left( x_{i5}+x_{j5}-x_{ij} \right) \quad  \,\, i,j=1,2,3,4.
\ee
The simplex inequalities are precisely determined by the following theorem: 
\begin{proposition}[$4$-Simplex Realizability (c.f. \cite{Deskter:1987,Asante:2021zzh,Shoenberg})]\label{Realizable}
A set of real numbers $\{x_{12},\cdots,x_{45} \}$   is realizable as the set of squared edge lengths of a 4-simplex $\Delta_4$ 
in 
$\mathbb R^4$ (flat Euclidean spacetime) or $\mathbb R^{1,3}$ (Minkowski spacetime) if and only if the corresponding Gram matrix $M_{\Delta_4}(x)$ is positive definite (signature $(+,+,+,+)$) or pseudo-definite (signature $(-,+,+,+)$), respectively. 
\end{proposition}
We call the length solutions that satisfy the Euclidean (resp.~Minkowski) realizability condition outlined in Proposition \ref{Realizable}, {\blue Euclidean} (resp.~{\blue Lorentzian}) {\blue geometric }solutions. Otherwise, they are termed {\blue non-geometric} solutions. Similarly, we refer to the corresonding simplices as {\blue geometric}/{\blue non-geometric} or {\blue Euclidean}/{\blue Lorentzian} simplices accordingly. 

\begin{table}[ht!]
\centering
\begin{tabular}{|C{4em}|C{5em}|C{4.5em}||C{5em}|C{4.5em}|}\hline
 \multirow{2}{4em}{\hspace{0.5cm}  Interval} & \multicolumn{2}{C{9.5em}||}{Lorentzian solutions} &  \multicolumn{2}{C{9.5em}|}{Euclidean solutions}  \\ \cline{2-5}
& Percentage of positive solutions  & Max \vspace{0.3cm} & Percentage of positive solutions& Max \vspace{0.3cm}\\ \hline 
  $I_{0.0}$  & $99.370\%$ & 25 & $0.1496\%$ & 5 \\ \hline
  $I_{0.3}$  & $98.5686\%$ & 29 & $1.4152\%$ & 5 \\ \hline
  $I_{0.5}$  & $96.7452\%$ & 26 & $3.2548\%$ & 5 \\ \hline
\end{tabular}
\caption{Statistics of Positive Length Solutions}
\label{tab:LorEcl}
\end{table}

We refine our experiment from the previous section by calculating the percentages of the positive solutions in our experiment which satisfy the Euclidean/Lorentzian constraints. This data is displayed in Figure \ref{fig:hist2} via the sub-figures \ref{fig:hist2A}, \ref{fig:hist2B}, and \ref{fig:hist2C}. In each interval, over {\blue 96\%} of the positive solutions satisfy the Lorentzian simplex inequalities as shown in Table \ref{tab:LorEcl}. Only a small fraction of the positive solutions satisfy the Euclidean simplex inequalities.  As in the previous section for real/positive solutions, we remark that as $\alpha \to 0.0$, the distribution of Lorentzian solutions seems to approach a Gaussian distribution.
{ Interestingly, however, the parameters \eqref{64sols} produce the maximum number of Lorentzian solutions, leading to the following corollary of Theorem \ref{thm:64Real}.
\begin{corollary}
\label{cor:32Lorentzian}
There exists integer areas so that there are $32$ Lorentzian solutions to the area-length system. 
\end{corollary}
Given Theorem \ref{thm:64solutionsnumerical}$^*$, Corollary \ref{cor:32Lorentzian} states that the maximum number of Lorentzian solutions possible, is indeed attainable, specifically by the integer areas \eqref{64sols}.
}
%Interestingly, all the $32$ positive solutions associated to the set of area parameters in \eqref{64sols} satisfy Lorentzian simplex inequalities. 

In Minkowski spacetime $\mathbb R^{3,1}$, the signature of squared volumes for the positive dimensional faces of a simplex determine their causal characters. By convention, a {\blue{spacelike}} triangle is characterized by a positive square area, a {\blue{timelike}} triangle has negative squared area, and a {\blue{lightlike}} triangle has vanishing squared area. One can allow for all possible signatures for the triangles in a $4$-simplex by specifying the appropriate values to the area parameters that determine its area geometry. Consequently, length geometries compatible with the specified area geometry can be determined through solutions to the area-length systems.

\begin{remark}
 A set of area parameters for a 4-simplex may result in a solution set that contains both Euclidean and Lorentzian simplices as viable solutions. Therefore, a set of ten area parameters assigned to the triangles of a 4-simplex apriori do not determine its causal character. However, Theorem \ref{thm:64solutionsnumerical}$^*$ still holds for all signatures of the triangle areas, and bounds the number of (isolated) realizations by $N=64$.
\end{remark}

{ 
We conclude our experiments by considering solutions to the area-length systems \eqref{AlSys2} for squared area parameters randomly selected from a normal distribution with mean 0 and standard deviation 1.  This distribution of parameters encompasses both positive squared areas (representing spacelike triangles) and negative squared areas (representing timelike triangles).  Note that, the presence of negative areas excludes the existence of Euclidean length solutions due to the constraints imposed by the simplex inequalities. Our results on an experiment with one-milion samples are displayed in Figure~\ref{fig:hist3}. Note the qualitatively distinct distributions of real, Lorentzian, and positive solutions in Figure~\ref{fig:hist3} as compared to Figures \ref{fig:hist} and \ref{fig:hist2}.

\begin{figure}[htb!]
\centering
\includegraphics[scale=0.45]{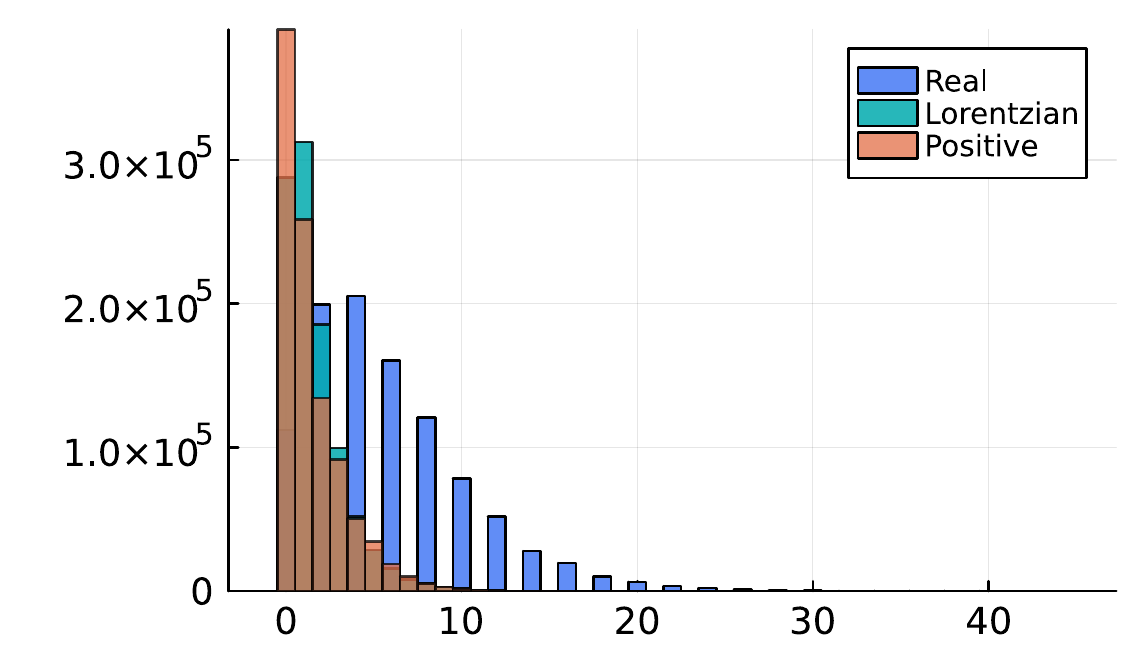}
\caption{Histograms for the counts of real, positive and Lorentzian solutions to the area-length system. The area parameters are selected from a normal distribution with mean 0 and standard deviation 1.}
\label{fig:hist3}
\end{figure}

}

\section{The Galois/Monodromy group and additional symmetry constraints}\label{sec:symmetry}

Equipped with the ability to track solutions of the area-length system via a parameter homotopy, it is computationally simple to track solutions over loops in the parameter space. Given a loop ${\blue{\gamma}}:[0,1] \to \mathbb{C}_{\textbf{v}}^{10}$  of generic values 
in the parameter space, based at some parameter $\gamma(0)=\gamma(1) = {\blue{\textbf{v}^{(1)}}}$, the set of solutions $S = \{s_1,\ldots,s_{64}\}~\subset ~\mathbb{C}_{\textbf{x}}^{10}$ to the polynomial system $F_{\Delta_4}(\textbf{x};\textbf{v}^{(1)})={\bf 0}$ 
permute via the bijection between start solutions and target solutions induced by the solution paths. In this way, every loop $\gamma$ in the parameter space  based at $\textbf{v}^{(1)}$ corresponds to a permutation ${\blue{\sigma_\gamma}}$ in the symmetric group ${\blue \mathfrak S_{64}}$. The collection of all permutations obtainable this way is called the {\blue monodromy group} or {\blue Galois group} of the area-length system. This group is well-defined up to relabelling the solutions in $S$ and does not depend on the generic base point chosen. We write ${\blue G} \subseteq \mathfrak S_{64}$ for this group. For more details about Galois/monodromy groups of enumerative problems, see \cite[Section 4]{NumericalNonlinearAlgebra} or \cite{Yahl}.

Symmetries in the solutions of a polynomial system restrict the possible permutations in the monodromy group. For example, direct inspection reveals that, for any $\textbf{v}^{(1)} \in \mathbb{C}_{\textbf{v}}^{10}$, the point  $\textbf{x}^*$ is a solution to $F_{\Delta_4}(\textbf{x};\textbf{v}^{(1)})$ if and only if $-\textbf{x}^*$ is a solution as well. As a consequence, we may label the solutions as 
\[S = \{s_1,\ldots,s_{32}\} \cup \{-s_1,\ldots,-s_{32}\}
\] and we see that 
\[\sigma_\gamma(s_i) = \sigma_{\gamma}(s_j) \iff \sigma_{\gamma}(-s_i) = \sigma_{\gamma}(-s_j).
\] This partition forces the group $G$ to be a subgroup of the {\blue wreath product} $\mathbb Z_{2} \wr \mathfrak S_{32}$ where the $32$ pairs may permute among each other freely, and the pairs themselves may also permute independently and freely. Hence the order of this group is $|G| = 32!\cdot 2^{32}$.

Explicit numerical calculations show that the containment $G \leq \mathbb{Z}_2 \wr \mathfrak S_{32}$ is an equality: By generating several loops and tracking the $64$ solutions along them, we produce permutations in $G$ which generate $\mathbb{Z}_2 \wr \mathfrak S_{32}$. This method is subject to numerical error and although, in theory, it can be certified (e.g. using certified path-tracking \cite{CertifiedPathTracking}), we do not perform this certification.

\begin{theorem}[*]\label{thm:monodromygroup}
The monodromy group $G$ of $F_{\Delta_4}$  is isomorphic to $\mathbb Z_{2} \wr \frak S_{32}$.
\end{theorem}

The geometric viewpoint of the group $G$, realized as the permutations of solutions to the area-length system as the parameters move, is what makes $G$ a {\textit{monodromy group}}. We also call $G$ a {\textit{Galois group}} because Harris in \cite{Harris} showed that monodromy groups of branched covers are Galois groups of some associated field extension. As a consequence, the existence of a formula (involving arithmetic operations $+,-,\times,\div$ and taking $n$-th roots) for the $64$  possible sets of edge lengths of a $4$-simplex in terms of generic triangular areas is exactly encoded by the \textit{solvability} of $G$. The group $G = \mathbb{Z}_2 \wr \mathfrak S_{32}$ is not solvable, and so we obtain the following corollary.

\begin{corollary}[*]
\label{cor:notsolvable}
There is no formula which expresses the edge lengths of a $4$-simplex in terms of the (generic) triangular areas that uses only the operations $+,-,\times, \div$ and $n$-th roots.
\end{corollary}

Corollary \ref{cor:notsolvable}$^*$ underlines the importance of numerical methods like homotopy continuation for studying the solutions to area-length systems. One way to recover hope for finding edge-length formulae in terms of triangular areas is to restrict the space of parameters to some subspace. Doing so restricts the number of loops that are possible in the parameter space, and thus, induces a smaller monodromy/Galois group, which may be solvable. 
We revisit this strategy in Section \ref{sec:Mono}.

\subsection{Symmetry conditions of the area-length System: 4-Simplex}

Consider the stratification of the $10$-dimensional space $\mathbb{C}_{\bf v}^{10}$ into strata based on which of the ten areas of a $4$-simplex are equal. Specifically, consider a  {\blue set partition} denoted by ${\blue{\nu}}=({\blue{\nu_1,\ldots,\nu_k}})$ of the $2$-dimensional faces of a $4$-simplex into ${\blue k}$ parts. We call the number partition ${\blue{|\nu|}}=(|\nu_1|,\ldots,|\nu_k|)$ of $10$ its {\blue signature}. Any such set partition indexes the set 
\[
{\blue Q_\nu} = \left\{{\bf v} \in \mathbb{C}_{\textbf{v}}^{10} \middle| \begin{array}{c} v_a=v_b  \text{ if and only if } \\ a,b \in \nu_i  \text{ for some } i\end{array}\right\}.
\]
Moreover, the parameter space $\mathbb{C}_{\textbf{v}}^{10}$ is a \textit{disjoint} union $\bigcup_{\nu} Q_\nu$ of the strata $Q_\nu$. Similarly, we have the stratification of the squared-edge space $\mathbb{C}_{\textbf{x}}^{10} = \bigcup_{\mu} P_\mu$ where $\mu$ ranges over all set partitions ${\blue{\mu}}=({\blue \mu_1,\ldots,\mu_k})$ of the edges of a $4$-simplex, and 
\[
{\blue P_\mu} = \left\{{\bf x} \in \mathbb{C}_{\textbf{x}}^{10} \middle| \begin{array}{c} x_a=x_b  \text{ if and only if } \\ a,b \in \mu_i  \text{ for some } i\end{array}\right\}.
\]
The closure $\overline{P_\mu}$ is isomorphic to $\mathbb{C}^{k}$. Moreover, this closure contains $P_{\mu'}$ if and only if $\mu$ is a refinement of $\mu'$. We remark that the closure $\overline{P_{\textbf{1}}} = \mathbb{C}_{\textbf{x}}^{10}$ contains every stratum (here ${\blue \textbf{1}}$ refers to the unique set partition with signature $(1,1,\ldots,1)$) and that $P_{\textbf{10}}$ is contained in the closure of every $P_\mu$ where ${\blue \textbf{10}}$ is the unique set partition of signature $(10)$. The analogous statements about the strata $Q_\nu$ in  $\mathbb{C}_{\textbf{v}}^{10}$ hold as well.

There are  $\blue B_{10} = 115\,975 $ (where $B_n$ is the $n$-th Bell number) many set partitions of the ten triangular faces (or edges) of a $4$-simplex, and hence, $115\,975$ many strata of $\mathbb{C}_{\textbf{v}}^{10}$ or $\mathbb{C}_{\textbf{x}}^{10}$ one may consider. Up to the symmetry induced by the action of $\mathfrak S_5$ on the vertices of $\Delta_4$, there are only $\blue B_{10}^s = 1299$, a number which can be computed using \texttt{OSCAR}'s group theory functionality provided by \texttt{GAP} \cite{OSCAR,GAP}.

So far, one may think of our results as pertaining to the largest stratum, $Q_{\textbf{1}}$. We will not extend our results to the area-length systems over each of the remaining $115\,974$ strata, however, the techniques in the previous sections can be directly applied to do so. Instead, we illustrate how the situation changes by considering the other extreme: the area-length system over parameters in the smallest stratum $Q_{\textbf{10}}$. After, we briefly discuss how the area-length system behaves when the edge-length solutions are constrained  to strata $P_\mu \subseteq \mathbb{C}_{\textbf{x}}^{10}$. 

Throughout, it is useful to refer to the following diagram. It shows the map ${\blue \varphi}:\mathbb{C}_{\textbf{x}}^{10} \to \mathbb{C}_{\textbf{v}}^{10}$ which sends a $10$-tuple of squared edge lengths of a $4$-simplex to its ten squared triangular areas. When constrained to a stratum $P_{\mu}$, the image of this map lies uniquely in some $\overline{Q_{\nu}}$, though it is not necessarily surjective.

\begin{figure}[ht!]
\centering 
\begin{tikzcd}[column sep=large,row sep=large]
  \mathbb C_{\bf x}^{10} \arrow[r, "\varphi"]  &  \mathbb C_{\bf v}^{10}  \\
  P_{\mu} \arrow[r, "\varphi |_{P_\mu}"] \arrow[u, hook]
& \overline{{ Q}_{\nu}}  \arrow[u, hook ] 
\end{tikzcd}
\caption{Stratification of the edge-length variables and area parameters of a 4-simplex. The map $\varphi$ takes ten edge lengths of a 4-simplex and outputs ten triangle areas using Heron's formula. }
\label{fig:SymmetryDiagram}
\end{figure}

\subsection{Symmetries of area parameters}\label{sec:Mono}
In this section we consider the restriction of the area-length system to some symmetry stratum  $Q_\nu$. As shown in previous sections, the monodromy group of the area-length system over $Q_{\textbf{1}}$ is $G = \mathbb{Z}_2 \wr \mathfrak S_{32}$. As we consider smaller strata, the monodromy group of the area-length system restricted to those strata produce subgroups of $G$. In other words, denoting $G_{\nu}$ to be the monodromy group of the area-length system restricted to $Q_{\nu}$, we have that $G_{\nu} \leq G_{\nu'}$ whenever $Q_{\nu} \subseteq \overline{Q_{\nu'}}$. As a consequence, the poset of set-partitions given by refinement corresponds to a poset of subgroups of $G = G_{\textbf{1}}$ given by inclusion.

\subsubsection{Example: Symmetry with equal area parameters}\label{sols:Equiarea}

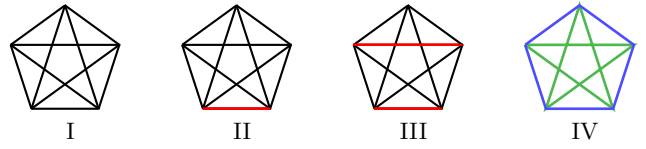
\begin{figure}[ht!]
\centering
\begin{tikzpicture}[scale=0.76,rotate=90]
  % Vertices
  \foreach \i in {1,...,5}
    \coordinate (V\i) at (72*\i:1);
    %\node[circle, fill, inner sep=0pt] (V\i) at (72*\i:1) {};

  % Edges
  \foreach \i in {1,...,5}{
    \foreach \j in {\i,...,5}{
      \draw[thick] (V\i) -- (V\j);
    }
  }  
  \node at (-1.2,-0.1) {${\rm I}$};
\begin{scope}[yshift=-3cm]
% Vertices
  \foreach \i in {1,...,5}
    \coordinate (V\i) at (72*\i:1);
    %\node[circle, fill, inner sep=0.7pt] (V\i) at (72*\i:1) {};
   \foreach \i in {1,...,5}{
   \draw[thick, red] (V2) -- (V3);
    \foreach \j in {\i,...,5}{
      \draw[thick] (V\i) -- (V\j);
    }
  }  
\node at (-1.2,-0.1) {${\rm II}$};
\end{scope}
\begin{scope}[yshift=-6cm]
% Vertices
  \foreach \i in {1,...,5}
    \coordinate (V\i) at (72*\i:1);
    %\node[circle, fill, inner sep=0.7pt] (V\i) at (72*\i:1) {};
   \foreach \i in {1,...,5}{
   \draw[thick, red] (V2) -- (V3) (V1) -- (V4);
    \foreach \j in {\i,...,5}{
      \draw[thick] (V\i) -- (V\j);
    }
  }  
\node at (-1.2,-0.1) {${\rm III}$};
\end{scope}
\begin{scope}[yshift=-9cm]
% Vertices
  \foreach \i in {1,...,5}
    \coordinate (V\i) at (72*\i:1);
    %\node[circle, fill, inner sep=0.7pt] (V\i) at (72*\i:1) {};

\draw[line width=1.0,green!60!black!70] (V1) -- (V3) -- (V5) -- (V2) -- (V4) -- (V1);
\draw[line width=1.0, blue!70] (V1) -- (V2) -- (V3) -- (V4) -- (V5) -- (V1);

\node at (-1.2,-0.1) {${\rm IV}$};
\end{scope}
\end{tikzpicture}
\caption{Distinct length solutions to area-length system of a 4-simplex with equal area parameters. Each coloured edge corresponds to a distinct edge length variable.}
\label{fig:A1}
\end{figure}

Consider the equi-area case $Q_{\textbf{10}} \cong \mathbb{C}^1_A$ where all triangles in a $4$-simplex share a common area $A$. The area-length system becomes
\begin{equation}\label{EquiArea}
F_{\Delta_4}({\bf x},A)\hspace{-1pt} = \hspace{-1pt}\left\{ 4 x_{ij}x_{ik} \hspace{-1pt}-\hspace{-1pt} (x_{ij} + x_{ik} - x_{jk})^2  \hspace{-1pt}- \hspace{-1pt}16A^2  \right\}
\end{equation}
for all $1 \leq\hspace{-2pt} i\hspace{-2pt}<\hspace{-2pt}j\hspace{-2pt}<\hspace{-2pt}k \hspace{-2pt}\leq 5$.
 For any area $A\in \mathbb C\backslash \{0\}$, each of the $32$ pairs $S = \{\pm s_i\}_{i=1}^{32}$ of solutions exhibits symmetry in one of four ways, that is, each belongs to one of four types of symmetry strata $P_{\mu}$.  By ordering the coordinates of the edges in  $\mathbb{C}_{\textbf{x}}^{10}$ as ${\bf x}~=~(x_{12},x_{13},x_{14},x_{15},x_{23},x_{24},x_{25},x_{34},x_{35},x_{45})$ and setting 
\begin{equation}\label{BetaGamma}
 {\blue{\beta}} = \frac{4\sqrt{A^2}}{\sqrt 3 }  , \q  \q {\blue{\gamma}} =  \frac{4\sqrt{A^2}}{\sqrt{5}} {\rm i},  
\end{equation}
the four types of solutions are as follows.
\begin{enumerate}
\item {Type $\textrm{I}$ (Equal Lengths):} The edges of the $4$-simplex  have equal lengths $\beta$. There is one pair of solutions of this type:
\begin{equation}\label{EquiA1}
 {\blue{S_{\rm I}}} = \{ s_1,-s_1  \}, \q \q  s_1 = (\beta, \beta, \ldots,\beta).   
\end{equation} 
In other words $S_{\rm I} \subseteq P_{\textbf{10}}$. 
For  $\beta\in \mathbb R_{>0}$, $s_1$ represents an equilateral 4-simplex, and hence it satisfies the Euclidean simplex inequalities. 
\item {Type ${\rm II}$ (of signature $(9,1)$):} Here, any fixed edge has length $3\beta$ and each of the remaining nine edges have length $\beta$. Hence, there are ten pairs of solutions of this type:
\begin{eqnarray}\label{EquiA2}
 {\blue{S_{\rm II}}} &=&  \{ s_2,\ldots, s_{11} \} \cup \{ -s_2,\ldots, -s_{11} \} , \q  \nn \\
\text{e.g.} \q  s_2 &=&  (3\beta, \beta, \ldots,\beta,\beta).  \q  \q 
\end{eqnarray}
For $\beta \in \mathbb R_{>0} $, the positive solutions $ \{ s_2,\ldots, s_{11} \}$ satisfy the Lorentzian simplex inequalities. In terms of the stratification of $\mathbb{C}_{\textbf{x}}^{10}$, there is one pair in each of the ten strata indexed by a set partition of signature $(9,1)$. 
\item {Type ${\rm III}$ (of signature $(8,2)$):} Here, any two fixed non-adjacent edges (see Figure \ref{fig:A1}) have lengths $3\beta$ and the remaining eight edges have length $\beta$. There are $15$ pairs of this type:
\begin{eqnarray}\label{EquiA2}
 {\blue{S_{\rm III} }}&=& \{ s_{12},\ldots, s_{26} \} \cup \{ -s_{12},\ldots, -s_{26} \} , \q  \nn \\
\text{e.g.} \q  s_{12} &=&  (3\beta, \beta, \ldots,\beta,3\beta).
\end{eqnarray}
For $\beta \in \mathbb R_{>0} $, the positive solutions $ \{ s_{12},\ldots, s_{26} \}$ also satisfy the Lorentzian simplex inequalities. 
\item {Type ${\rm IV} $ (Cyclic of signature $(5,5)$):} The remaining $6$ pairs
\[  {\blue{S_{\rm IV}}} = \{ s_{27},\ldots, s_{32} \} \cup \{ -s_{27},\ldots, -s_{32} \}   \]   of length solutions may be written in terms of $\gamma$.  
The coordinates  of these solutions $\{x_{ij}\}_{1 \leq i < j\leq 5}$ satisfy
\begin{eqnarray}
x_{ij} &=&  x_{jk}= x_{kl} = x_{lm} = x_{mi} = \gamma , \q   \nn \\
x_{ik} &=&  x_{km}= x_{mj} = x_{jl} = x_{li} = -\gamma    \\
\text{e.g.} \q  s_{27} &=&  (\gamma, -\gamma,-\gamma,\gamma,\gamma,-\gamma,-\gamma,\gamma,-\gamma,\gamma). \nn
\end{eqnarray}
Each solution has $5$ edges (forming a cycle) with length $\gamma$, and the remaining $5$ edges have length $-\gamma$. Note that for $A\in \mathbb R\backslash \{0\}$, $\gamma$ is pure imaginary, and these solutions are non-real.   
\end{enumerate}

The length geometries for the four types of solutions to the system \eqref{EquiArea} are depicted pictorially in Figure \ref{fig:A1}. 

Figure \ref{fig:A1} illustrates how $\varphi^{-1}(Q_{\textbf{10}})$ consists of $32$ pairs of solutions, appearing  in $32$ strata $P_\mu$ for $32$ distinct set-partitions $\mu$. These set partitions appear in four orbits under the action of $\mathfrak S_5$ on the vertices of $\Delta_4$, which we have termed Types $\rm{I}-\rm{IV}$. Their signatures are $(10), (9,1), (8,2),$ and $(5,5)$, although not every set-partition with those signatures appears (e.g. the $(8,2)$-signature partition requires that the edges in a part of size two do not meet at a vertex). We remark that $\varphi^{-1}(Q_\textbf{10})$ is \textit{not} equal to the union of these strata. To see this, note that $(2,1,\ldots,1,2)$ is not a possible solution, but belongs to the same stratum as $(3,1,\ldots,1,3)$ which is a possible solution. 

Nonetheless, for a generic point in ${\bf v} \in Q_{\textbf{10}}$, we have that $\varphi^{-1}({\bf v})$ has $ {\blue{k_{\mu}}}=2$ points in $P_{\mu}$ for $32$ distinct $\mu$. Writing $ {\blue{m_{\mu}}}$ for the size of the orbit of $\mu$ under the action of $\mathfrak S_5$ which stabilizes the area symmetry $Q_{{\bf 10}}$, we have
\begin{align}
64 = 2+\cdots+2 &= \hspace{-4pt}\sum_{\mu} k_{\mu}\\ &=\hspace{-5pt}\hspace{-7pt} \sum_{\text{Orbit}(\mu)} \hspace{-5pt}\hspace{-1pt} k_{\mu} \hspace{-3pt}\cdot\hspace{-1pt} m_{\mu} = 2\hspace{-1pt}\cdot\hspace{-1pt} 1 + 2 \hspace{-1pt}\cdot\hspace{-1pt}  10 + 2 \hspace{-1pt}\cdot\hspace{-1pt}  15+ 2 \hspace{-1pt}\cdot\hspace{-1pt}  6 \nn
\end{align}
\begin{remark}
Since over the smallest stratum $Q_{{\bf 10}}$, there are generically $64$ distinct simple solutions, all strata have this property. Consequently, there is some formula similar to the one above which partitions the $64$ solutions over a generic point in $Q_{\nu}$ depending on their membership in some $P_{\mu}$. We leave the problem of determining these formulae for all $1299$ orbits of strata to future research.
\end{remark}

By the way we have written each of the $64$ solutions explicitly (and in terms of radicals) it is clear that the monodromy group $G_{\textbf{10}}$ of the area-length system, restricted to the stratum $Q_{\textbf{10}}$ is isomorphic to $\mathbb{Z}_2$: the only permutations possible are induced by following $A \in \mathbb{R}-\{0\}$ in a loop around $0$, thus permuting $\beta \leftrightarrow -\beta$ and $\gamma \leftrightarrow -\gamma$.

\begin{remark}
As remarked upon before, we leave it open to determine $G_{\nu}$ for other partitions $\nu$ of the triangular faces of $\Delta_4$. Similarly, we leave it open to determine the maximal such partitions which induce a solvable group, and the minimal such partitions which induce a transitive group.
\end{remark}

\subsection{Symmetries of length variables}\label{sec:lengthsymmetry}

Since the edge-lengths of a triangle determine its area via Heron's formula, we immediately see that edge-length symmetries described by $P_\mu$ induce area symmetries described by some $Q_\nu$. Specifically,  given any set partition $\mu$ of the edges of $\Delta_4$ we have that $\varphi(P_\mu) \subseteq \overline{Q_\nu}$ for some set-partition $\nu$ of the triangular faces of $\Delta_4$ such that $\overline{Q_{\nu}}$ is minimal with this property. 
In fact, $\nu$ is the set partition describing the area symmetries of a simplex with generic squared edge lengths ${\textbf{x}} \in P_\mu$.  

\begin{figure}[!htpb]
\begin{tikzpicture}[scale=0.76,rotate=90]
\begin{scope}[scale=2]
  \foreach \i in {1,...,5}
    \coordinate (V\i) at (72*\i:1);
    %\node[circle, fill, inner sep=0.7pt] (V\i) at (72*\i:1) {};
   \foreach \i in {1,...,5}{
   \draw[thick, red] (V2) -- (V3) (V2) -- (V1);
    \foreach \j in {\i,...,5}{
      \draw[thick] (V\i) -- (V\j);
    }
  }  
\node at (-1.2,-0.1) {};
\end{scope}

\begin{scope}[yshift=-4cm]
% Vertices
  \foreach \i in {1,...,5}
    \coordinate (V\i) at (72*\i:1);
    %\node[circle, fill, inner sep=0.7pt] (V\i) at (72*\i:1) {};
   \foreach \i in {1,...,5}{
   \draw[thick, red] (V2) -- (V3) (V2) -- (V1);
   \draw[thick, black] (V1) -- (V3) ;
    \foreach \j in {\i,...,5}{
      \draw[thick,gray] (V\i) -- (V\j);
    }
  }  
\node at (-1.2,-0.1) {};
\end{scope}
\begin{scope}[yshift=-4cm,xshift=2.2cm]
% Vertices
  \foreach \i in {1,...,5}
    \coordinate (V\i) at (72*\i:1);
    %\node[circle, fill, inner sep=0.7pt] (V\i) at (72*\i:1) {};
   \foreach \i in {1,...,5}{
   \draw[thick, black]  (V3) -- (V4);
   \draw[thick, black] (V1) -- (V4) (V1) -- (V3);
    \foreach \j in {\i,...,5}{
      \draw[thick, gray] (V\i) -- (V\j);
    }
  }  
\node at (-1.2,-0.1) {};
\end{scope}
\begin{scope}[yshift=-4cm,xshift=-2.2cm]
% Vertices
  \foreach \i in {1,...,5}
    \coordinate (V\i) at (72*\i:1);
    %\node[circle, fill, inner sep=0.7pt] (V\i) at (72*\i:1) {};
   \foreach \i in {1,...,5}{
   \draw[thick, red]  (V2) -- (V3);
   \draw[thick, black] (V2) -- (V5) (V5) -- (V3);
    \foreach \j in {\i,...,5}{
      \draw[thick, gray] (V\i) -- (V\j);
    }
  }  
\node at (-1.2,-0.1) {};
\end{scope}
\end{tikzpicture}
\caption{A depiction of an edge-symmetry stratum $P_{\mu}$ (depicted on left) which, under $\varphi$, does not surject onto its area-symmetry stratum $Q_{\nu}$ (depicted on right).}
\label{fig:nonsurjectivestratum}
\end{figure}
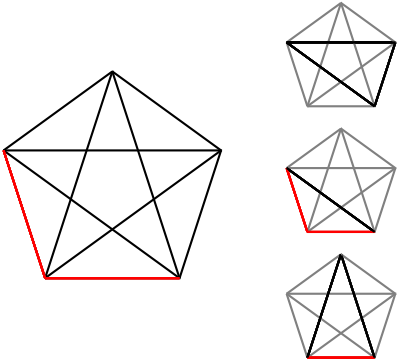

Determining $\nu$ from $\mu$ is combinatorial and straightforward. Triangles of $\Delta_4$ indexed by triples $(i,j,k)$ and $(i',j',k')$ belong to the same part of $\nu$ if and only if the edges $\{ij,ik,jk\}$ and $\{i'j',i'k',j'k'\}$ belong to the same parts of $\mu$. For example, the edge-partition
\[
\mu = (\{12,23\},\{13,\ldots,45\})
\]
describes a $2$-dimensional stratum $P_\mu$. It is illustrated combinatorially in Figure \ref{fig:nonsurjectivestratum}. For a generic element $\textbf{x}^*~\in~P_\mu$, the areas $\varphi(\textbf{x}^*)$ exhibit the symmetry imposed by the partition
\[
\nu = (\{134,135,145,245,345\},\{124,125,234,235\},\{123\})
\]
on the triangles of $\Delta_4$. Since $Q_\nu$ is $3$-dimensional, the map $\varphi$ restricted to $P_{\mu}$ cannot be surjective. In addition to the symmetry constraints described by $Q_{\nu}$, the squared areas must additionally satisfy the formula
\[
3v_{123}^2-14v_{123}v_{345}+16v_{235}v_{345} - 5v_{345}^2=0.
\]

A partition $\nu$ of triangles in $\Delta_4$ does not determine a unique partition $\mu$ of the edges: given a generic point ${\bf v}  \in Q_\nu$, the preimage $\varphi^{-1}({\bf v})$ may consist of $10$-tuples of squared edge-lengths which showcase several distinct symmetry patterns. This was shown in the previous section, for example, in the case of $\nu = \textbf{(10)}$.

\section{Area-Length systems for general triangulations}\label{sec:Triangulations}

The relationship between area and length geometries extends beyond the scope of a 4-simplex. Here, we shall illustrate two  strategies to solve the area-length systems within four-dimensional triangulations, composed of multiple 4-simplices.  These triangulations generally consist of interconnected 4-simplices through shared 3-simplices (tetrahedra). Similar to the case of a single 4-simplex, a triangulation's area geometry is described by assigning area values to its constituent triangles.

Let $\blue{\cal T}$ denote a triangulation with $\blue{N_\Delta}$ $4$-simplices, $\blue{N_t}$ triangles, and $\blue{N_e}$ edges. Its area-length system, expressed using Heron's formula \eqref{TrgsHeron}, is given by
\begin{eqnarray}\label{TrgsHeron}
\hspace{-1pt} {\blue{F_{\cal T}({\bf x}\,;{\bf v})}} \hspace{-2pt}=\hspace{-2pt} \left\{4 x_{ij}x_{ik} \hspace{-2pt}-\hspace{-2pt} (x_{ij} + x_{ik} - x_{jk})^2  \hspace{-2pt}- \hspace{-1pt}16v_{ijk}
\right\}_{t_{ijk}} 
\end{eqnarray}
where the indices range over all triangles ${\blue t_{ijk}}$ in the triangulation $\mathcal T$, $x_{ij},x_{ik},x_{jk}$ are the squared edge lengths of triangle $t_{ijk}$, and $v_{ijk}$ is its square area. For fixed areas, this system consists of $N_t$ quadratic equations in the squared edge lengths. Over the complex numbers, the variables ${\bf x} \in \mathbb C_{\bf x}^{N_e} $ denote the set of squared edge length variables, while ${\bf v} \in \mathbb C_{\bf v}^{N_t}$ represents the set of area parameters.

In a generic four-dimensional triangulation, the number of triangles $N_t$ exceeds that of edges $N_e$. For instance, a non-trivial triangulation of any closed compact $4$-manifold satisfies the inequality $N_t \geq \tfrac{4}{3}N_e$  \cite{Barrett:1997tx}, leading to an {\blue over-determined} area-length system.  Moreover,  the difficulty of solving \eqref{TrgsHeron} increases with the number of simplices involved.

Although the homotopy continuation methods discussed in Section \ref{sec:Homotopy} can be directly applied to solve the over-determined system \eqref{TrgsHeron}, doing so in this direct fashion requires tracking $2^{N_e}$ (the B\'ezout bound for the system) many paths in the total degree homotopy \eqref{totaldegreehomotopy}. The number $2^{N_e}$ quickly becomes prohibitively expensive, even for triangulations of moderate size. 

The system \eqref{TrgsHeron} can be alternatively addressed by leveraging the solutions to the square area-length systems localized on each $4$-simplex within the triangulation. 
Specifically, given a triangulation $\mathcal T$ involving the $4$-simplices $\{{\blue{\Delta^{(1)},\ldots,\Delta^{(N_{\Delta})}}}\}$, we write $\blue{\bf x}^{(i)}$ and $\blue{\bf v}^{(i)}$ for the squared edge lengths and squared triangle areas in $\Delta^{(i)}$ and consider the mapping
\begin{eqnarray}
{\blue \rho} : \mathbb C_{ \bf x}^{N_e}  & \longrightarrow &  \mathbb C_{ {\bf x}^{(1)}}^{10}  \times  \mathbb C_{ {\bf x}^{(2)}}^{10}  \times \ldots \times \mathbb C_{ {\bf x}^{(N_\Delta)}}^{10}   \nn\\
{\bf x}  & \mapsto & \left( {\bf x}^{(1)} , {\bf x}^{(2)}  \ldots , {\bf x}^{(N_\Delta)} \right). \q \q \q 
\end{eqnarray}
The image ${\blue{X_{\mathcal T}}}\subseteq  \mathbb C_{ {\bf x}^{(1)}}^{10}  \times \ldots \times \mathbb C_{ {\bf x}^{(N_\Delta)}}^{10}$ of $\rho$ is in bijection with $\mathbb{C}_{\textbf{x}}^{N_e}$. The problem of determining whether a point $\textbf{x}^*=({\bf{x}^*}^{(1)},\ldots,{\bf{x}^*}^{(N_{\Delta})} )\in  (\mathbb C_{ {\bf x}}^{10})^{(N_\Delta)} $ is in $X_{\mathcal T}$ is simple. The condition is that for all $i,j$,
\begin{equation} \label{TetConstraint}
{\hspace{-5pt}}{\bf x^*}^{(i)}_{e} = {\bf x^*}^{(j)}_{e} ,\quad  \text{for each edge } e \subseteq  \Delta^{(i)} \cap  \Delta^{(j)}.
\end{equation}
We refer to \eqref{TetConstraint} as the {\blue{length-matching conditions}}. 

Consider the collection of the local area-length systems of all $4$-simplices in $\mathcal T$:  
\begin{equation}\label{ALTrg2}
{\blue {\cal F}_{\Delta} ({\bf x}\,;{\bf v}) }= \left\{  F_{\Delta_4^{(i)}} ( {\bf x}^{(i)}\,;{\bf v}^{(i)} ) \right\}_{1\leq i \leq {N_\Delta} }.
\end{equation} These $N_{\Delta}$-many square systems can be efficiently solved  via the parameter homotopy method \eqref{eq:parameterhomotopy}, resulting in the $N_{\Delta}$ many local solution sets $\blue{S^{(1)},\ldots,S^{(N_{\Delta})}}$. In total, there are $64\cdot N_{\Delta}$ solutions.  Determining which combinations ${\textbf{s}} \in S^{(1)}\times \ldots \times S^{(N_{\Delta})}$  combine to provide a solution to the system \eqref{TrgsHeron} is easily done by checking the length-matching conditions \eqref{TetConstraint} for all overlapping $4$-simplices on the candidate $\textbf{s}$;  this is the same as checking if $\textbf{s}$ belongs to $X_{\mathcal T}$.

Note that the area-length system \eqref{TrgsHeron} for a triangulation remains invariant under the transformation ${\bf x} \mapsto - {\bf x}$, therefore,  if ${ \bf s} $ solves the system, then so does $-{ \bf s} $.

\begin{remark}
For a large triangulation with generic area parameters, satisfying all of the  conditions of the form  \eqref{TetConstraint} becomes challenging, resulting in limited or no length solutions to its area-length system. 

The alternate strategy for solving the area-length systems is crucial for two reasons: Firstly, it is a more efficient way of finding solutions to the over-determined system \eqref{TrgsHeron}, since the square systems for the 4-simplices can be solved efficiently in parallel. In addition, it is relatively fast and easy to implement the length matching conditions \eqref{TetConstraint}. Secondly, in the construction of spin foam models for quantum gravity, a weak form of the conditions \eqref{TetConstraint} is implemented. Therefore, the solutions to the area-length systems localized on 4-simplices within a triangulation become valuable. See the discussion in Section \ref{sec:ESFM} for more details.
\end{remark}

\subsection{Area-length system for triangulation comprising two 4-simplices: Equi-area case}

Let's apply the alternate strategy to the first non-trivial triangulation ${\blue{T_2}} $: two  4-simplices $\Delta_4^{(1)}$ and $\Delta_4^{(2)}$  which intersect along a $3$-simplex. Table \ref{tab:T2} summarizes some important numbers related to $T_2$. 

\begin{table}[ht!]
\centering
\begin{tabular}{|C{3em}|C{3em}|C{3em}|}\hline
 $N_\Delta $ &  $N_t$ & $N_e$  \\ \hline
 $2$ &  $16$ & $14$   \\ \hline 
\end{tabular}
\caption{Number of components of  triangulation $T_2$.}
\label{tab:T2}
\end{table}
The corresponding area-length system $F_{T_2}({\bf x},{\bf v})$ is therefore over-determined, as it consists of $16$ equations in $14$ variables. After ``squaring up'', the B\'ezout number associated to $F_{T_2}(\textbf{x},\textbf{v})$ is  $2^{14} = 16384$ and for generic areas, it has no solutions.
Hence, we consider a simpler example case, where all the area-parameters are set equal to $A$, so that  ${\bf v}^* = \left( A^2 , A^2, \ldots, A^2\right) \in \mathbb C_{\bf v}^{16}$.

The 64 length solutions to the area-length system of each 4-simplex are grouped into four distinct types according to signatures of their length geometries: 
\begin{equation}\label{EqSols}
S^{(i)} =  S_{\rm I}^{(i)}  \cup S_{\rm II}^{(i)}  \cup S_{\rm III}^{(i)}  \cup S_{\rm IV}^{(i)} , \q     \q  i= 1,2.
\end{equation}
These solutions are explicitly described in Section \ref{sols:Equiarea} using $\beta, \gamma$ expressed in terms of $A$ in Equation \eqref{BetaGamma}.

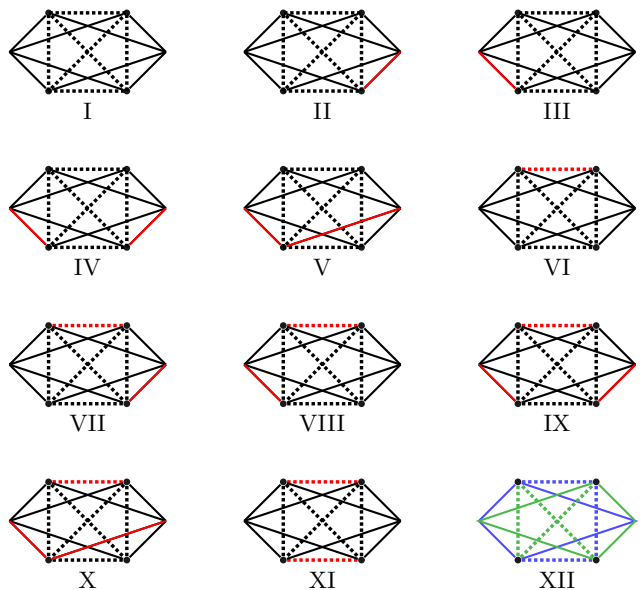
\begin{figure}[ht!]
\centering
\begin{tikzpicture}[scale=0.52]

\node[circle, fill, black!90, inner sep=0.9pt] (v1) at (0,0) {};
\node[circle, fill, black!90, inner sep=0.9pt]  (v2) at (0,2) {};
\node[circle, fill, black!90, inner sep=0.9pt]  (v3) at (2,0) {};
\node[circle, fill, black!90, inner sep=0.9pt]  (v4) at (2,2) {};

\draw[densely dotted, line width=1.3] (v1) rectangle (v4) (v1)--(v4) (v2)--(v3);

\foreach \i in {1,...,4}{
\draw[thick] (-1,1) --(v\i);
\draw[thick] (3,1) --(v\i);
}
\node at (1,-0.5) {$\rm I$};%{$S_{\rm I} $-$S_{\rm I} : \tau_A $};

%\node[circle, fill, black!20!purple!70, inner sep=1.5pt] (v3)  {};

\begin{scope}[xshift = 6cm ]
\node[circle, fill, black!90, inner sep=0.9pt] (v1) at (0,0) {};
\node[circle, fill, black!90, inner sep=0.9pt]  (v2) at (0,2) {};
\node[circle, fill, black!90, inner sep=0.9pt]  (v3) at (2,0) {};
\node[circle, fill, black!90, inner sep=0.9pt]  (v4) at (2,2) {};

\draw[densely dotted, line width=1.3] (v1) rectangle (v4) (v1)--(v4) (v2)--(v3);

\foreach \i in {1,...,4}{
\draw[thick] (-1,1) --(v\i);
\draw[thick] (3,1) --(v\i);
} 
\draw[thick,red] (3,1) --(v3);

\node at (1,-0.5)  {$\rm II$};%{${S}_{\rm I} $-$ {S}_{\rm II}: \tau_A $};

\end{scope}
\begin{scope}[xshift = 12cm ]
\node[circle, fill, black!90, inner sep=0.9pt] (v1) at (0,0) {};
\node[circle, fill, black!90, inner sep=0.9pt]  (v2) at (0,2) {};
\node[circle, fill, black!90, inner sep=0.9pt]  (v3) at (2,0) {};
\node[circle, fill, black!90, inner sep=0.9pt]  (v4) at (2,2) {};

\draw[densely dotted, line width=1.3] (v1) rectangle (v4) (v1)--(v4) (v2)--(v3);

\foreach \i in {1,...,4}{
\draw[thick] (-1,1) --(v\i);
\draw[thick] (3,1) --(v\i);
} 
\draw[thick,red] (-1,1) --(v1);

\node at (1,-0.5)  {$\rm III$};%{${S}_{\rm I} $-$ {S}_{\rm II}: \tau_A $};

\end{scope}

\begin{scope}[yshift = -4cm ]
\node[circle, fill, black!90, inner sep=0.9pt] (v1) at (0,0) {};
\node[circle, fill, black!90, inner sep=0.9pt]  (v2) at (0,2) {};
\node[circle, fill, black!90, inner sep=0.9pt]  (v3) at (2,0) {};
\node[circle, fill, black!90, inner sep=0.9pt]  (v4) at (2,2) {};

\draw[densely dotted, line width=1.3] (v1) rectangle (v4) (v1)--(v4) (v2)--(v3);

\foreach \i in {1,...,4}{
\draw[thick] (-1,1) --(v\i);
\draw[thick] (3,1) --(v\i);
}
\draw[thick,red] (-1,1) --(v1) (3,1) --(v3);
\node at (1,-0.5)  {$\rm IV$};%{${S}_{\rm II} $-$ {S}_{\rm II}: \tau_A $};
\end{scope}

\begin{scope}[yshift = -4cm, xshift = 6cm  ]
\node[circle, fill, black!90, inner sep=0.9pt] (v1) at (0,0) {};
\node[circle, fill, black!90, inner sep=0.9pt]  (v2) at (0,2) {};
\node[circle, fill, black!90, inner sep=0.9pt]  (v3) at (2,0) {};
\node[circle, fill, black!90, inner sep=0.9pt]  (v4) at (2,2) {};

\draw[densely dotted, line width=1.3] (v1) rectangle (v4) (v1)--(v4) (v2)--(v3);

\foreach \i in {1,...,4}{
\draw[thick] (-1,1) --(v\i);
\draw[thick] (3,1) --(v\i);
}
\draw[thick,red] (-1,1) --(v1)--(3,1);
\node at (1,-0.5)  {$\rm V$};%{${S}_{\rm II} $-$ {S}_{\rm II}: \tau_A $};
\end{scope}

\begin{scope}[yshift = -4cm, xshift = 12cm ]
\node[circle, fill, black!90, inner sep=0.9pt] (v1) at (0,0) {};
\node[circle, fill, black!90, inner sep=0.9pt]  (v2) at (0,2) {};
\node[circle, fill, black!90, inner sep=0.9pt]  (v3) at (2,0) {};
\node[circle, fill, black!90, inner sep=0.9pt]  (v4) at (2,2) {};

\draw[densely dotted, line width=1.3] (v1) -- (v2) -- (v3)--(v4) --(v1)--(v3);
\draw[densely dotted, line width=1.3,red] (v2)--(v4);

\foreach \i in {1,...,4}{
\draw[thick] (-1,1) --(v\i);
\draw[thick] (3,1) --(v\i);
}
\node at (1,-0.5)  {$\rm VI$};%{${S}_{\rm II}$-$ {S}_{\rm II}: \tau_B $};
\end{scope}

\begin{scope}[yshift = -8cm ]
\node[circle, fill, black!90, inner sep=0.9pt] (v1) at (0,0) {};
\node[circle, fill, black!90, inner sep=0.9pt]  (v2) at (0,2) {};
\node[circle, fill, black!90, inner sep=0.9pt]  (v3) at (2,0) {};
\node[circle, fill, black!90, inner sep=0.9pt]  (v4) at (2,2) {};

\draw[densely dotted, line width=1.3] (v1) -- (v2) -- (v3)--(v4) --(v1)--(v3);
\draw[densely dotted, line width=1.3,red] (v2)--(v4);

\foreach \i in {1,...,4}{
\draw[thick] (-1,1) --(v\i);
\draw[thick] (3,1) --(v\i);
}
\draw[thick,red] (3,1) --(v3);
\node at (1,-0.5)  {$\rm VII$};%{${S}_{\rm II} $-$ {S}_{\rm III} : \tau_B $};
\end{scope}

\begin{scope}[yshift = -8cm, xshift = 6cm ]
\node[circle, fill, black!90, inner sep=0.9pt] (v1) at (0,0) {};
\node[circle, fill, black!90, inner sep=0.9pt]  (v2) at (0,2) {};
\node[circle, fill, black!90, inner sep=0.9pt]  (v3) at (2,0) {};
\node[circle, fill, black!90, inner sep=0.9pt]  (v4) at (2,2) {};

\draw[densely dotted, line width=1.3] (v1) -- (v2) -- (v3)--(v4) --(v1)--(v3);
\draw[densely dotted, line width=1.3,red] (v2)--(v4);

\foreach \i in {1,...,4}{
\draw[thick] (-1,1) --(v\i);
\draw[thick] (3,1) --(v\i);
}
\draw[thick,red] (-1,1) --(v1);
\node at (1,-0.5)  {$\rm VIII$};%{${S}_{\rm II} $-$ {S}_{\rm III} : \tau_B $};
\end{scope}

\begin{scope}[yshift = -8cm, xshift = 12cm ]
\node[circle, fill, black!90, inner sep=0.9pt] (v1) at (0,0) {};
\node[circle, fill, black!90, inner sep=0.9pt]  (v2) at (0,2) {};
\node[circle, fill, black!90, inner sep=0.9pt]  (v3) at (2,0) {};
\node[circle, fill, black!90, inner sep=0.9pt]  (v4) at (2,2) {};

\draw[densely dotted, line width=1.3] (v1) -- (v2) -- (v3)--(v4) --(v1)--(v3);
\draw[densely dotted, line width=1.3,red] (v2)--(v4);

\foreach \i in {1,...,4}{
\draw[thick] (-1,1) --(v\i);
\draw[thick] (3,1) --(v\i);
}
\draw[thick,red] (3,1) --(v3) (-1,1)--(v1);
\node at (1,-0.5)  {$\rm IX$};%{${S}_{\rm III} $-$ {S}_{\rm III} : \tau_B $};
\end{scope}

\begin{scope}[yshift = -12cm ]
\node[circle, fill, black!90, inner sep=0.9pt] (v1) at (0,0) {};
\node[circle, fill, black!90, inner sep=0.9pt]  (v2) at (0,2) {};
\node[circle, fill, black!90, inner sep=0.9pt]  (v3) at (2,0) {};
\node[circle, fill, black!90, inner sep=0.9pt]  (v4) at (2,2) {};

\draw[densely dotted, line width=1.3] (v1) -- (v2) -- (v3)--(v4) --(v1)--(v3);
\draw[densely dotted, line width=1.3,red] (v2)--(v4);

\foreach \i in {1,...,4}{
\draw[thick] (-1,1) --(v\i);
\draw[thick] (3,1) --(v\i);
}
\draw[thick,red] (3,1) --(v1) (-1,1)--(v1);
\node at (1,-0.5)  {$\rm X$}; 
\end{scope}

\begin{scope}[yshift = -12cm , xshift = 6cm]
\node[circle, fill, black!90, inner sep=0.9pt] (v1) at (0,0) {};
\node[circle, fill, black!90, inner sep=0.9pt]  (v2) at (0,2) {};
\node[circle, fill, black!90, inner sep=0.9pt]  (v3) at (2,0) {};
\node[circle, fill, black!90, inner sep=0.9pt]  (v4) at (2,2) {};

\draw[densely dotted, line width=1.3] (v1) -- (v2) -- (v3)--(v4) --(v1);
\draw[densely dotted, line width=1.3,red] (v2)--(v4) (v1)--(v3);

\foreach \i in {1,...,4}{
\draw[thick] (-1,1) --(v\i);
\draw[thick] (3,1) --(v\i);
}
\node at (1,-0.5)  {$\rm XI$};%{${S}_{\rm III} $-$ {S}_{\rm III} : \tau_C $};
\end{scope}

\begin{scope}[yshift = -12cm, xshift = 12cm ]
\node[circle, fill, black!90, inner sep=0.9pt] (v1) at (0,0) {};
\node[circle, fill, black!90, inner sep=0.9pt]  (v2) at (0,2) {};
\node[circle, fill, black!90, inner sep=0.9pt]  (v3) at (2,0) {};
\node[circle, fill, black!90, inner sep=0.9pt]  (v4) at (2,2) {};

%\draw[thick,blue] (v1) rectangle (v4) (v1)--(v4) (v2)--(v3);

\draw[line width=1.3, densely dotted, blue!70] (v1)--(v3)--(v4)--(v2)  ;
\draw[thick, blue!70]  (v1)--(-1,1) -- (v2)   (v2)--(3,1)--(v1);
\draw[line width=1.3, densely dotted, green!60!black!70] (v3)--(v2)--(v1)--(v4) ;
\draw[thick,green!60!black!70] (v4)--(-1,1) -- (v3)   (v3)--(3,1)--(v4);

\node at (1,-0.5)  {$\rm XII$};%{${S}_{\rm IV} $-$ {S}_{\rm IV} : \tau_D $};
\end{scope}

\end{tikzpicture}
\caption{Length geometries of distinct length solutions to the area-length system for the triangulation $T_2$. }
\label{fig:T2}
\end{figure}

Out of the $\blue  64^2=4096$ possible combinations of the length solutions
only $\blue  176$ of them satisfy the length matching conditions \eqref{TetConstraint} and are in $X_{T_2}$. Hence,  the system $F_{T_2}$ for equal area parameters has 176 isolated length solutions which come in pairs as 
\[  \{ {\bf s}_i , -{\bf s}_i \}_{1\leq i \leq 88} \q \,\, \text{where } \,\, {\bf s}_i|_{\Delta^{(j)}} \in S^{(j)} . \]    The solutions are grouped into twelve distinct types, as depicted in Figure \ref{fig:T2}, according to length geometries of the 4-simplices and the shared tetrahedron. The edge lengths of the shared tetrahedron are represented by dotted lines in Figure \ref{fig:T2} and the counts of the solutions within each distinct type are summarized in Table \ref{tab:typesT2}. 

\begin{table}[ht!]
\centering
\begin{tabular}{|C{5.7em}||C{2.7em}|C{2.7em}|C{2.7em}|C{2.7em}|C{2.7em}|C{2.7em}|}\hline
Type & $\rm I$ & $\rm II$ & $\rm III$ & $\rm IV$ & $\rm V$ & $\rm VI$  \\ \hline
Frequency & $2$ & $8$ & $8$ & $24$ & $8$& $12$  \\
Type of $\Delta^{(1)}$ & \rm I&\rm I&\rm II&\rm II&\rm II&\rm II \\
Type of $\Delta^{(2)}$ & \rm I&\rm II&\rm I&\rm II&\rm II&\rm II \\
\hline 
\end{tabular}
%\vspace{0.5cm}
\begin{tabular}{|C{5.7em}||C{2.7em}|C{2.7em}|C{2.7em}|C{2.7em}|C{2.7em}|C{2.7em}|}\hline
Type & $\rm VII$ & $\rm VIII$ & $\rm IX$ & $\rm X$ & $\rm XI$ & $\rm XII$  \\ \hline
Frequency & $24$ & $24$ & $24$ & $24$ & $6$& $12$   \\
Type of $\Delta^{(1)}$ & \rm II&\rm III&\rm III&\rm III&\rm III&\rm IV \\
Type of $\Delta^{(2)}$ & \rm III&\rm II&\rm III&\rm III&\rm III&\rm IV \\
\hline 
\end{tabular}

\caption{Number of solutions to the system $F_{T_2}$.}
\label{tab:typesT2}
\end{table}

\begin{remark}
The number of solutions in any of the above types equals twice the number of ways to draw a representative diagram like those in Figure \ref{fig:T2}. For example, there are $24$ solutions of type $\rm VII$ because producing one involves first picking one of six edges in $\Delta_4^{(1)} \cap \Delta_4^{(2)}$ and subsequently picking one of two non-intersecting edges of $\Delta_4^{(2)}$. For any such choice, there is a unique pair $(\textbf{s},-\textbf{s})$ of that type. 
\end{remark}

%\vspace{0.05cm}

\section{Applications: Length solutions from area variables }\label{sec:App}

In this section, we explore practical applications of homotopy continuation methods within the context of discrete gravity and spin foam models. The algorithm in Figure \ref{Jcode} facilitates the systematic computation of length solutions from the area parameters of a 4-simplex. These solutions can be utilized to generate sets of boundary data for a spin-foam vertex amplitude (refer to Appendix \ref{Coherent_data} for details on this construction). Additionally, the existence of multiple length solutions corresponding to generic area parameters of a 4-simplex allows us to quantify the ambiguities involved in defining geometric quantities associated with the 4-simplex in terms of the area variables. Such ambiguities have implications in area Regge calculus and effective spin foam models.

\subsection{Geometric quantities from area variables}

Focusing on a 4-simplex, its area-length system produces up to $64$ solutions (refer to Theorem \ref{thm:64solutionsnumerical}$^*$)
for generic area parameters. These solutions split into two sectors, denoted as `plus' and `minus', represented by
\[ S= {\blue S^+} \cup {\blue S^-} = \{s_1,  \ldots, s_{32} \} \cup \{-s_1,  \ldots, -s_{32} \} ,  \] 
where these sectors are reflections of each other. Concentrating on a specific sector (e.g., the plus sector $S^+$),  there may exist up to 32 edge length assignments for defining geometric quantities for a 4-simplex from its prescribed ten triangle areas. One approach to address these ambiguities is by constraining the edge lengths to specific symmetries, potentially leading to a unique length solution. Subsection \ref{sec:lengthsymmetry} discusses implications on imposing symmetry conditions on the edge lengths through set partitions into equal values. Once a unique length solution is obtained, geometric quantities like volumes, angles, etc., associated with faces within the 4-simplex can be deduced from well-known formulae. Such symmetry reductions are proposed in \cite{Barrett:1997tx}, aiming to resolve ambiguities in defining the action for area Regge calculus. However, these symmetry conditions may be overly restrictive, particularly for larger triangulations with generic area parameters.

Alternatively, the multiple length solutions associated to a 4-simplex can be combined to define geometric quantities from its area variables. We propose an averaging over all compatible length solutions to the corresponding area-length system. For example, let $\{ s_1,\ldots, s_k \}$ be the set of length geometries (for e.g. subset of solutions in the sector $S^+$ satisfying simplex inequalities) compatible with a given set of area parameters $\bm a$. Then, a geometric quantity $V$ can be defined in terms of the area variables as
\begin{equation}\label{def:areageometry}
V(\{ \bm a \}) := \frac1k\sum_{i=1}^{k}  V(\{ s_i \} ) \q 
\end{equation}  
by averaging over the corresponding geometric quantity expressed  in terms of the edge lengths. 
{ This definition is natural when the quantity being computed is a sum over the 4-simplices and the length solutions are considered in the systems for the individual 4-simplices.  }

Geometric quantities related to area variables, expressed in terms of multiple length variables, also appear in {\textit{area-metric formulations}} \cite{Schuller:2005yt}. {\blue{Area-metrics}} are conceptualized as algebraic curvature maps defined on the space of anti-symmetric bi-vectors. Thus, they allow for measurements of areas of surfaces, hence, are closely related to the area variables employed here. For further details on area-metric geometries, refer to \cite{Schuller:2005yt} for continuous manifolds, and adapted in \cite{Dittrich:2023ava} for discrete simplicial manifolds.

In the subsequent subsection, we provide a summary of the key features of effective spin foam models and shed light on the role played by the multiple length solutions for 4-simplices in defining their amplitudes.

\subsection{Effective spin foam models}\label{sec:ESFM}

Effective spin foam models ({\blue ESFM}), as explored in \cite{Asante:2020qpa,Asante:2020iwm,Asante:2021zzh}, constitute a family of discrete gravitational path integral formulations that utilize area variables as their fundamental quantities on a fixed triangulation. These models encode the discrete area spectrum inherent in quantum geometries derived from loop quantum gravity, assigning (almost) equidistantly spaced eigenvalues to the areas of triangles. 

The amplitudes for the effective spin foam models are defined through a summing over discrete area variables and expressed as 
\begin{equation}\label{eqn:ESFs}
	{\blue {Z}_{\rm ESFM}} = \sum_{\{{\bm a} \}} \mu({\bm a})\,  e^{\text{i} S_{\rm ARC}({\bm a})} \, \prod_\tau G_\tau ({\bm a}) .
\end{equation}
Here, ${\blue \mu({\bm a})}$ is a measure term\footnote{ The measure terms are yet to be determined. They can be fixed, for instance,  by demanding discretization invariance.} for the area variables. $\blue S_{\rm ARC}$ represents the area Regge action for the given triangulation and its exponential term in \eqref{eqn:ESFs} captures the oscillatory behaviour of the amplitudes. The $\blue G_\tau$ represent Gaussian terms that implement constraints between area variables of neighbouring 4-simplices. The solutions to the area-length system for 4-simplices play a pivotal role in defining both the area Regge action and the Gaussian terms within the ESFM amplitudes. 

A key aspect of ESFM is given by the incorporation of the so-called {\blue area-length constraints}. These constraints, localized onto bulk tetrahedra, ensure consistent length assignments derived from the area variables of neighbouring 4-simplices and are implemented through the Gaussian terms $G_\tau$. The length-matching conditions, as outlined in \eqref{TetConstraint} for pairs of 4-simplices, precisely match the area-length constraints associated with a tetrahedron $\tau$. 

It is worth noting that the six edge lengths of a tetrahedron can be uniquely (up to reflection) inverted to its four triangle areas and two dihedral angles at adjacent edges (See \cite{Asante:2018wqy}, Appendix C). For a tetrahedron $\tau$,  this establishes a one-to-one correspondence: 
\begin{align*}  
 \{   x_{ij}  \}_{1\leq i<j\leq 4}  \,\,  \leftrightarrow  \,\,   \{  v_{123}, v_{124}, v_{134}, v_{234}, \phi_{12}^\tau, \phi_{13}^\tau  \} 
 \end{align*}
where $v_{ijk}$ is the (squared) area of triangle ${(ijk)}$ and ${\blue \phi_{ij}^\tau}$ is the dihedral angle at the edge ${(ij)}$. 

Making use of this correspondence, the conditions \eqref{TetConstraint}  can be transformed into conditions between the triangle areas and dihedral angles of the shared tetrahedron. Given that the areas are parameters, the triangular faces of the shared tetrahedron possess identical areas in the neighbouring 4-simplices.  Consequently, the constraints for the four areas are automatically satisfied. Thus, the length-matching conditions reduce to the following two conditions \cite{Asante:2020qpa}: 
\begin{equation}\label{TetConstraint2}
 \phi_{e}^{\tau (i)} = \phi_{e}^{\tau(j)} , \q e \in \{ 1,2\} . 
\end{equation}
Here, $ \phi_{e}^{\tau (i)} $ represent the dihedral angles of two adjacent edges of tetrahedron $\tau$ within the 4-simplex $\Delta_4^{(i)}$. Equations \eqref{TetConstraint2} are usually referred to as {\blue shape-matching constraints}. 

The operator form of these shape-matching constraints exhibits non-commutativity  \cite{KapvichM,Asante:2021zzh}, indicating an anomaly (parametrized by the Barbero-Immirzi parameter). Consequently, the constraints are of ``second-class" nature and are therefore imposed weakly, albeit as strongly as allowed by the anomaly, via the Gaussian terms $G_\tau$ in the amplitudes \eqref{eqn:ESFs}.

For a generic set of area parameters, utilizing the algorithm  {in Figure \ref{Jcode}} for each 4-simplex produces multiple solutions (potentially even when constrained by the simplex inequalities) for the length variables. As a consequence, the shape-matching conditions \eqref{TetConstraint2} occur many times.  This multitude of constraints contribute to the ESFM amplitudes through products of the Gaussian terms $G_\tau$, which weakly enforce them. Moreover, the multiple length solutions for each 4-simplex also contributes to the area Regge action in \eqref{eqn:ESFs}.

The area Regge action, expressed as a sum over triangles and 4-simplices within the triangulation $\cal T$ is given by 
\begin{equation}\label{ARaction}
S_{\rm ARC}({\bm a}) =  \sum_{t \in \cal T}  n_t \, a_t   - \sum_{\Delta \in \cal T } \sum_{t \supset \Delta} a_t \,\theta_t ({\bm a}), \,\,
\end{equation}
where ${\blue \theta_t}$ is the dihedral angle hinged at triangle $t$ with area ${\blue a_t}$ inside a 4-simplex $\Delta$, and ${\blue n_t}$ represents the total angle (half the angle if $t$ is on the boundary) around a point in the 2D subspace containing the triangle. 
This action quantifies `curvature' distribution across the triangulation. The dihedral angles inside a 4-simplex as a function of area variables are typically inferred from inversion of areas and lengths (refer to \cite{Asante:2021zzh} for dihedral angles in terms of edge lengths). Therefore, the multiple length solutions for generic area parameters contribute to $S_{\rm ARC}$ through a sum over the dihedral angles in \eqref{ARaction}. 

{ 
The area variables utilized within ESFM amplitudes \eqref{eqn:ESFs} assume a discrete set of values which are (almost) equidistantly spaced. These discrete areas stem from the spectrum of area operators derived from LQG  \cite{Rovelli:1994ge}. Specifically, the areas assigned to space-like triangles take the form $ a_t \sim \ell_P^2 \gamma j $, where $ \ell_P = \sqrt{8\pi \hbar G/c^3} $ denotes Planck length, $ \gamma $ represents the Barbero-Immirzi parameter which parametrizes the area gap, and $ j \in \mathbb{N}/2 $ are half-integers. Time-like triangles, on the other hand, are assigned area values of the form $ a_t \sim \ell_P^2 n $, where $n \in \mathbb{N} $ are positive integers.
For these discrete area variables, the area-length system for a 4-simplex maintains the same symmetries as previously discussed. Thus, Theorems \ref{thm:64solutionsnumerical}$^*$ and \ref{thm:64Real} hold generically for these configurations as a consequence of the discussion in Remark \ref{rem:rationals}.  

The nature of the physically relevant length solutions  for these discrete area configurations requires further detailed analysis and study. However, we expect that in general there will be multiple Lorentzian solutions that contribute to the amplitudes \eqref{eqn:ESFs}. This is so, even in the simplest, symmetric case where all the areas are equal: there is one Euclidean solution but multiple Lorentzian solutions (as discussed in Section \ref{sols:Equiarea}). The results in Section \ref{sec:Results} suggest that if the variation of the area parameters is large (i.e. large area gap), then there are slightly fewer physically relevant solutions compared to area parameters with small variations in the area parameters.

Of interest is the asymptotic behaviour of the amplitude $Z_{\rm ESFM}$ with respect to large areas. Previous works \cite{Asante:2020qpa,Asante:2020iwm} reproduced length Regge calculus dynamics by analyzing expectation values of certain geometric observables for varying boundary areas. As the amplitudes \eqref{eqn:ESFs} depend on the solutions to the area-length system, a first step would be an analysis of the asymptotics of the solutions themselves. Such an analysis belongs to the field of \textit{tropical geometry} \cite{Tropical} where one would study the \textit{tropical solutions} of the area-length system (see similar techniques in \cite{Agostini}).  We leave this to future work.
}

In summary, the homotopy methods implemented {in Figure \ref{Jcode}} prove very useful for computations of the effective spin foam amplitudes for generic area parameters.

\section{Discussion}

Various quantum gravity approaches, such as LQG, spin foam models, and holography, use area variables that characterize a class of `non-metric geometries' as their fundamental degrees of freedom. To ensure consistency with classical general relativity (GR) in the appropriate limit, it is important to establish a connection between these area variables and the length variables inherent in GR. 

Area Regge Calculus (ARC) is a discrete formulation of gravity that employs areas assigned to triangles to describe the geometry of a piecewise flat simplicial complex.  Each fundamental cell, given by a 4-simplex, is comprised of ten edges and ten triangles, thus one may infer the geometry of a  flat 4-simplex solely from its triangle areas. 
Effective spin foam models (ESFM) approach to quantum gravity utilizes area Regge calculus in its formulation and also provides a mechanism to control the mismatch between geometries from neighbouring 4-simplices, particularly at their shared tetrahedra with prescribed areas to the triangular faces. Allowing this mismatch in a controlled manner turns out to be a key motivation for using area variables. Thus, understanding the nature of area-length systems is of central importance. The ESFM approach, therefore, poses a mathematical challenge of understanding the area-length system, which is addressed in this article through the application of homotopy continuation method of numerical algebraic geometry.

Homotopy continuation methods from numerical algebraic geometry provide very useful tools to analyze and make several inferences about these area-length systems. We have performed extensive computer experiments for the area-length systems of a 4-simplex utilizing the algorithm {in Figure \ref{Jcode}}, and observed that generically there are $64$ (complex) isolated solutions for the squared edge-length variables given the ten triangle areas of the 4-simplex as parameters. Using the \texttt{certify} function within \texttt{HomotopyContinuation.jl}, we obtained a proof that there are at least $64$ isolated complex solutions to a generic instance of the area-length system for a $4$-simplex. Theorem \ref{Thm:Solutions}$^*$ claims that this number is exactly $64$.  Moreover, by tracking loops in the area parameter space that subsequently permute the solutions, we found the corresponding Galois group  (Theorem \ref{thm:monodromygroup}$^*$) of this enumerative problem. Since that Galois group is {\it not} solvable, we obtained Corollary \ref{cor:notsolvable}$^*$ which implies the absence of formulae for the lengths as functions of the area parameters in terms of radicals. However, imposing certain symmetries on the area parameter space can lead to solvability in terms of basic arithmetic operations.

While instances exist where certain area parameters lead to $64$ real solutions, this is not common; generally, the number of real solutions is much less than $64$. Furthermore, not all the $64$ solutions correspond to (length) geometric 4-simplices. In general, there are significantly fewer solutions that satisfy Euclidean 4-simplex inequalities compared to Lorentzian 4-simplex inequalities. { These physically relevant solutions are derived from semialgebraic conditions on the 64 complex length solutions.} These results support the understanding that area variables describe an enlarged space of geometries compared to the length or metric variables \cite{Dittrich:2008ar,Dittrich:2010ey,Freidel:2010aq}.

The study of the area-length systems within triangulations is relevant  for spin foam models, particularly, for effective spin foam models (ESFM). In these models, the dynamics governing discrete, independent area variables are determined by imposing shape-matching constraints among them. It has been demonstrated that for relative small, symmetry-reduced triangulations, the weak imposition of the constraints implements the dynamics of discrete gravity within a certain range of allowed parameters \cite{Asante:2020qpa,Asante:2020iwm}. The length solutions of the area-length systems obtained through numerical homotopy methods provide explicit expressions for these constraints within the dynamics of ESFM. Notably, the constraints, localized on shared tetrahedron of neighbouring 4-simplices, appear multiple times for generic area variables (including discrete areas) without imposing any symmetries. This is  also present in area Regge action, and can be incorporated into the ESFM amplitudes. 

For larger triangulations, the homotopy continuation methods offer fast and reliable approaches to determine numerical approximate solutions to the area-length systems. One can simultaneously solve for all the length solutions associated with  discrete area variables assigned locally to 4-simplices. Hence, the numerical homotopy methods employed within the effective spin foam models will facilitate in the study of refinement limit of spin foam models.

\appendix

\section{Algorithm to Solve the Area-Length System of a 4-Simplex} \label{app:Jcode}

In this section, we present an algorithm to solve the area-length system of a 4-simplex. The algorithm leverages the prototype \texttt{julia} package \texttt{Pandora.jl}, whose numerical calculations are handled by   \texttt{HomotopyContinuation.jl}, package for solving the associated polynomial equations.  The code for many of the computations performed in this article is provided below in Figure \ref{Jcode}. 

\begin{figure}[ht!]
\centering
\includegraphics[scale=0.5]{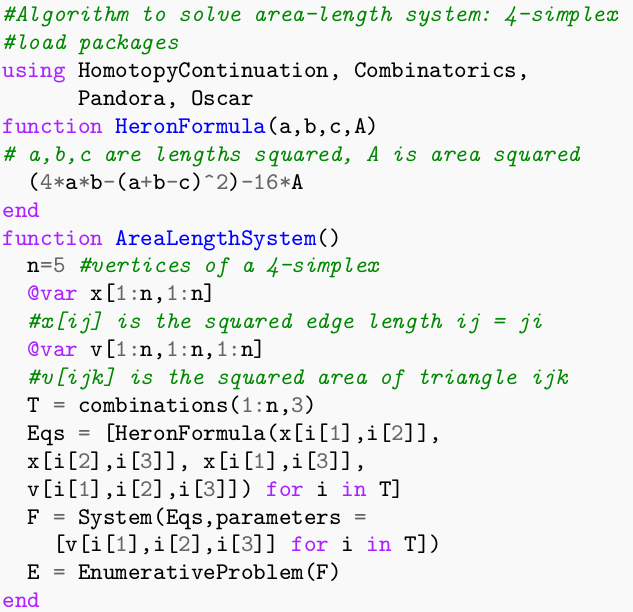} \\ \vspace{10pt}
\includegraphics[scale=0.5]{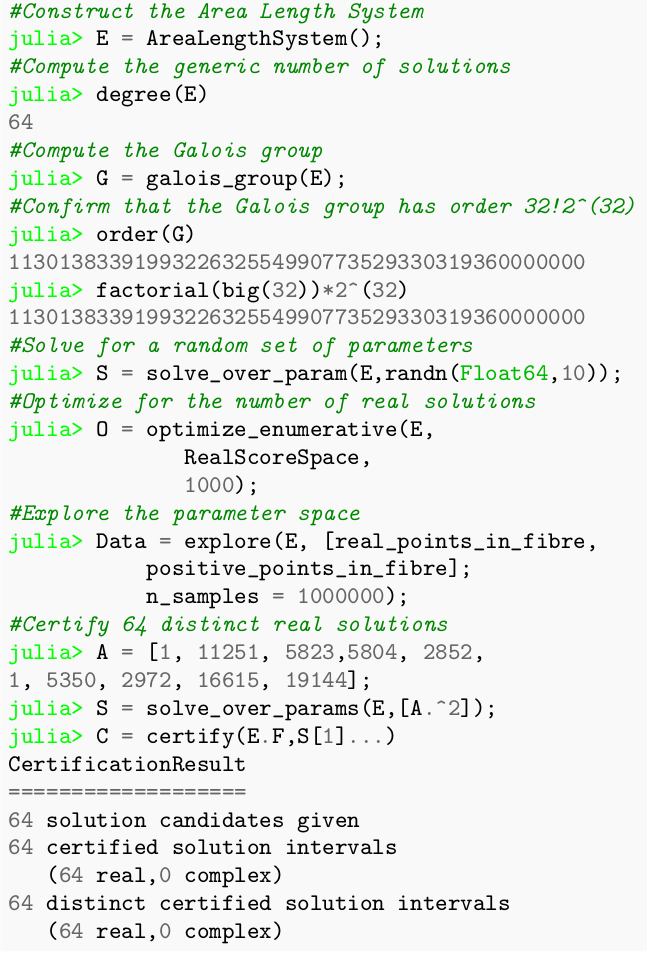}
\caption{Sample of a \texttt{julia} code using homotopy continuation, through \texttt{Pandora.jl}, to compute several properties of the area-length system of a 4-simplex.}
\label{Jcode}
\end{figure}

%\begin{listing}[H]
%\centering
%\inputminted[escapeinside=||,bgcolor=gray!2]{julia}{Heron4Simplex2.jl}
%\caption{sds}
%\label{Jcode}
%\end{listing}

The algorithm starts by loading the necessary packages. It proceeds to apply the Heron formula, computing the area of a triangle in terms of its squared edge lengths. Following this, it constructs the area-length system of equations for a 4-simplex, using squared areas as parameters and squared edge lengths as variables. This system is realized as an \texttt{EnumerativeProblem}, the main datatype of \texttt{Pandora.jl}. We compute the generic number of solutions to this system using the \texttt{degree} function, and proceed to compute its Galois group, verifying that it has order $32!\cdot 2^{32}$. Additionally, we provide code to solve this system over random parameter values, to explore the parameter space of the problem, and to attempt to optimize for the number of real solutions.

\section{Constructing boundary data for a coherent vertex in spin foam models}\label{Coherent_data}

Spin foam models constitute gravitational state sum models over quantum geometries defined on a 2-complex graph, often dual to a triangulation. In the coherent representation \cite{Livine:2007vk} { (for Euclidean signature)}, the boundary data are characterized by an over-complete basis consisting of spins $j_t$ and coherent intertwiners $ h_\tau $. The coherent representation allows for a geometric description in the semi-classical limit of these models. The spins $j_t$, label unitary irreducible representations of the underlying gauge group $\cal G$ and are associated with faces of the 2-complex. Geometrically, these spins represent triangle areas within the dual triangulation. Coherent intertwiners $h_\tau$,  associated with the edges (dual to a tetrahedron) of the 2-complex, consist of 3D unit normal vectors $\vec n_t \in S^2$ assigned to the triangular faces of the tetrahedron. { For Lorentzian signature, the unit normal vectors associated with the triangles lie either on a one-sheeted hyperbola $H_{\rm sp}$ if the triangle is spacelike or a two-sheeted hyperbola $H_\pm$ if the triangle is timelike. }

\begin{figure}[ht!]
\centering
\begin{tikzpicture}[scale=0.82,rotate=90]
%\draw[red,fill=red] (0,0) circle (.7ex); 
  % Vertices
  \foreach \i in {1,...,5}
    %\coordinate (V\i) at (72*\i:1);
    \node[rectangle, fill, inner sep=1.7pt] (V\i) at (72*\i:1) {};

  % Edges
  \foreach \i in {1,...,5}{ 
    \foreach \j in {\i,...,5}{
      \draw[thick] (V\i) -- (V\j);
      
    } %\draw[red,thick] (0,0) -- (V\i);
  }  

\node[yshift=2.5mm ,xshift=-4mm] at (V1) {\intertwiner{60} }; 
\node[xshift=-3mm,yshift =-4.2mm ] at (V2) {\intertwiner{140} }; 

\node[xshift=4mm, yshift = -4.2mm ] at (V3) {\intertwiner{220} }; 

\node[xshift=5mm,yshift=2.8mm] at (V4) {\intertwiner{300} }; 

\node[xshift=0.5mm, yshift=4mm] at (V5) {\intertwiner{0} }; 

%\node[xshift=14.5mm, yshift=4.5mm] at (V4) {$\{ \vec n_i\}_{1\leq i \leq 4}$};

\end{tikzpicture}
\caption{Representation of a coherent spin foam vertex. Lines represent spins $j_t$ assigned to faces of $\Gamma_v$ and the normal vectors $\vec n_i$ assigned to the faces of tetrahedra are denoted by the open circles. }
\label{fig:vertex}
\end{figure}
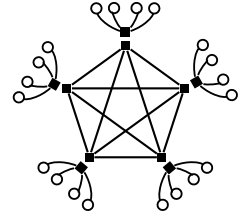

The simplest 2-complex is given by a vertex denoted by $\Gamma_v$. In four dimensional spin foam models, the vertex is $5$-valent and dual to a 4-simplex triangulation. The boundary data associated to dual triangulation consist of ten triangle areas and twenty unit normal vectors associated to the triangular faces of the tetrahedra (see Figure \ref{fig:vertex}). 

The algorithm detailed in Figure \ref{Jcode} efficiently computes all compatible length geometries corresponding to a given set of ten triangle areas. Subsequently, from these length solutions, several geometric quantities associated to the 4-simplex can  be determined.
These geometric quantities typically have explicit formulas in terms of its edge lengths.  For instance, the components of normal vectors, associated with the triangular faces of each tetrahedron, can be derived from the cofactor of its corresponding 3D Gram matrix (whose components are defined  in \eqref{GramMatrix}). These normal vectors, together with the triangle areas, provide the boundary data for computing coherent spin foam amplitudes. 
Thus for a fixed set of spins of $\Gamma_v$, the algorithm {in Figure \ref{Jcode}} allows for the identification of all geometric boundary data compatible with the boundary spins, commonly referred to as Regge boundary data. 

{ The proposed method for constructing boundary data for a spin foam vertex amplitude outlined here may play a significant role in comparing ESFMs with traditional spin foam models such as, the EPRL model \cite{Engle:2007wy} or the Conrady-Hnybida model \cite{Conrady:2010kc}, by studying the properties of the resulting amplitudes. 
}

\begin{acknowledgments}
SKA is supported by the Alexander von Humboldt Foundation. TB is supported by NSERC Discovery grant (RGPIN-2023-03551). The authors are grateful to John Barrett, Bianca Dittrich, Sebastian Steinhaus for helpful discussions. { We also thank the reviewers for their very helpful comments.}
\end{acknowledgments}

\bibliography{AreaLength}

\end{document}